%% file: HMS-CoinTossX-001.tex
\documentclass[final,5p,twocolumn]{elsarticle}
\include{Preamble}

\begin{document}
%------------------------------------------------------
%	FRONT MATTER
%------------------------------------------------------
\begin{frontmatter}
    \title{Simulation and estimation of a point-process market-model with a matching engine}
    
    \author[a1]{Ivan Jericevich}
    \ead{jrciva001@myuct.ac.za}
    \author[a1]{Patrick Chang}
    \ead{chnpat005@myuct.ac.za}
    \author[a1]{Tim Gebbie}
    \ead{tim.gebbie@uct.ac.za}
    
    \address[a1]{Department of Statistical Sciences, University of Cape Town, Rondebosch 7700, South Africa}
    
    \begin{abstract}
     The extent to which a matching engine can cloud the modelling of underlying order submission and management processes in a financial market and the importance of this with regards to the statistical robustness of models and resulting inference remains an unanswered concern for market models. Here we consider a 10-variate Hawkes process with simple rules to simulate common order types which are submitted to a matching engine. Hawkes processes can be used to model the time and order of events, and how these events relate to each other. However, they provide a freedom with regards to implementation mechanics relating to the prices and volumes of injected orders. This allows us to consider a reference Hawkes model and two additional models which have rules that change the behaviour of limit orders. The resulting trade and quote data from the simulations are then calibrated and compared with the original order generating process to determine the extent with which implementation rules can distort model parameters. Evidence from validation and hypothesis tests suggest that the true model specification can be significantly distorted by market mechanics to the extent that the models are no longer statistically equivalent. It can be argued that the matching mechanism merely blurs the estimation and identification of the process and this is a triviality. However, we argue that this can lead to fundamental changes in the nature of decisions when using these types of models. 
    \end{abstract}
\end{frontmatter}

\tableofcontents

%------------------------------------------------------
%	INTRODUCTION
%------------------------------------------------------
\section{Introduction \label{sec:Introduction}}

A matching engine is the key piece of technology that intermediates agents submitting orders with the aggregation of orders and then the emergence of prices, correlations and order book dynamics. The extent to which the matching engine can cloud the modelling of the underlying order submission and management processes remains an unanswered concern with regards to market models. Here we consider order submission models in the context of an experimental order matching engine -- CoinTossX \cite{CoinTossArxivPaper, sing2017cointossx}. The matching engine can provide a realistic platform for market model exploration and interaction. It may allow traders, organisations and academic institutions to test market structure, fragility and dynamics without the cost of live test trading, but with the realism of interaction via asynchronous market rules. 

The matching engine is independent of the modelling framework and order submission, and it directly exposes modelling to the vagaries of time-delays, message implementation rules and event driven asynchronicity. This implies that there can be a layer of inter-mediation that resides between the models, the data generation and observation processes whose impact on model decisions needs to be understood in terms of potential impacts on model identification, stability and interpretability.

This work demonstrates some aspects of this problem. Specifically the impact of the additional layer of complexity from the matching engine. There are two key aspects in this problem. First, given a model, can the model be reasonably identified and calibrated once model events are inject into a matching engine and then emerge from the data feeds? Second, how different can the identified model be from the actual model, or some reference model, when there are other practical considerations that come into play? In addition, this type of work can address some aspects of the unavailability of data and direct data feed access from industry to researchers, by providing a framework that can be compared to recorded transaction data arising from the actual market system interfaces using synthetic data \cite{RamanLiednerPMAS2019}. However, the type of model used here can at best be used to inform thinking about infrastructure integrity under various stress test scenarios as it cannot recover all the required stylised facts. In particular, a power law dependency in the limit order placement and other aspects of measured price impact \cite{LFM2003}. The intention of this work is not to code for stylised facts but to demonstrate how implementation rules can interact with model specifications in a simulated environment.  

The information content and dynamics of the limit order book (LOB) in order-driven markets have been studied extensively over the years with various stochastic, agent-based and Markov models being the most common representations 
% include bacry et al, Toki et al an abergel paper, and "Agent-based modeling: Methods and techniques for simulating human systems Eric Bonabeau PNAS www.pnas.org/cgi/doi/10.1073/pnas.082080899" ordered by date in the reference chain
\cite{BMM2015,Bonabeau2002,CTPA2011b,farmer2005predictive, zheng2014modelling}. These studies point to a number of important challenges associated with modelling an electronic limit order book. The first concerns the assumptions made -- in particular, whether order-flow exists due to the actions of ``perfectly rational'' (scrutinised for their inconsistency with direct observation of individual traders), or ``zero-intelligence'' traders or agents. The latter approach assumes order-flow is a result of aggregate behaviour that can be specified by a stochastic process. As a starting point, the ``zero-intelligence'' approach shows appeal for leading to easily quantifiable models that can yield falsifiable predictions \cite{gould2013limit}. A second challenge regards capturing key features relating to the feedbacks and nonlinear couplings that naturally occur in a market ecosystem mediated by an electronic limit order book.

Traders' actions depend on the state of the LOB and the state of the LOB also clearly depends on traders' actions. The estimation and identification of statistical properties with LOB data remains problematic -- several properties are observed to have power-law distributions {\it e.g.} order volumes, relative limit prices or the LOB depth profile, but there is no clear consensus on the best fitting procedure to adopt, nor the generation processes involved. Related to this is the stylised-fact centric validation of models with high-frequency data that ignore the problems of: i.) degenerate model parameters, and ii.) the inherently asynchronous and event driven nature of market clearing when modelling market in continuous trading \cite{PlattGebbie2016}. Other challenges include: the complexity of a LOB's state-space, hidden liquidity, bilateral trade agreements and opening/closing auctions \cite{gould2013limit}.

One of the first attempts at modelling the LOB in financial markets came from \citet{stigler1964public}. The approach taken at the time was to constrain prices within 10 ticks and randomly draw bid or ask orders (with equal probability) with prices uniformly distributed on the price grid. Each time an order crosses the opposite best quote, it becomes a market order. All orders are of size one. Orders not executed $N = 25$ time steps after their submission are cancelled. Thus, $N$ is the maximum number of orders available in the order book.\footnote{Other early models of the continuous double auction include \cite{garman1976market, gode1993allocative, gould2013limit, mike2008empirical}.}

A ``zero intelligence'' model of the LOB was implemented by \citet{farmer2005predictive} showing that constraints imposed by market institutions can often dominate strategic agent behaviour. They test a simple model with minimally intelligent agents who place orders to trade at random --- essentially dropping agent rationality altogether --- which is demonstrated to produce good quantitative predictions. More specifically, two types of agents place order randomly in both price (limit orders only) and time (both market and limit orders). Patient agents submit limit orders that arrive according to the same Poisson rate and where the price is uniformly distributed in $-\infty < p_t < a_t$ for buy orders and $b_t < p_t < \infty$ for sell orders (where $p_t$, $a_t$, and $b_t$ are the price, best ask, and best bid respectively on the log-scale). Impatient agents submit market orders according to a constant Poisson rate. The sizes of market and limit orders are kept constant. Rates of buying and selling are kept equal. Based on this, order arrivals will alter the best and in turn change the boundary conditions of the limit order price distributions. Despite its simplicity the model predicts the spread and average price diffusion rate well by dropping strategic choice and only considering the constraints imposed by the continuous double auction. The importance of these findings lies in the simple laws relating prices to order flow. 

In reality agents behave strategically, but these findings suggest that there are often circumstances where agent behaviour is dominated by considerations other than strategy. It then seems reasonable to approach agent-based modelling by first gaining a good understanding of market institutions using minimally intelligent agents and then work upwards towards more strategic agents. The basic approach taken by \citet{farmer2005predictive} can be considered similar to the one that will be taken here, with the key difference being that we use a multivariate Hawkes process to generate order types and their timings.

More recently, \citet{zheng2014modelling} used a multivariate marked Hawkes process for the joint spread-price dynamics. This is made possible by incorporating a constraint on intensities that preserves the natural ordering of best bid and ask prices. Four event types are considered to model the best bid or ask moving up or down by one tick. The constraint preventing the downward (upward) movement of the best ask (bid) comes into effect when the spread is one tick. 

\citet{BMM2015} provide an extensive list of applications for point processes in finance which gives valuable insights into the price, spread and top-of-book dynamics. The key problem with an order-flow Hawkes model in a matching engine environment is that the arrival rates and the types of events do not depend on the state of the limit order book -- specifically relating new orders to the existing orders in order book and the spread. This is part of the motivation for the use of a minimally intelligent simulation model to better understand how Hawkes process modelled events can be distorted by realistic matching rules.

The aim of this paper is to determine the extent to which market mechanisms and practical considerations from the limit order book can distort standard Hawkes model parameters. This is accomplished by specifying two minimally intelligent models for the submission of orders to a matching engine. Section \ref{sec:Simulation} starts with a summary of the theory behind Hawkes processes together with a discussion on the simulation set-up performed with CoinTossX. This includes a discussion around the event-types considered, the rules for submitting events/orders, caveats of our simulations with CoinTossX, model parameter choice, order volume choice, and lastly basic results generated from each model implemented with CoinTossX. Section \ref{sec:Estimation} demonstrates the estimation procedure followed together with results informing the extent to which simulation model parameters are distorted. The maximum likelihood calibration routine and the resulting parameter estimates are compared through a number of qualitative and quantitative tests for goodness-of-fit. Section \ref{sec:discussion} performs a likelihood-ratio test to confirm our results and the implications are discussed. Finally, section \ref{sec:Conclusion} summarises our findings. 
% For detailed results regarding calibrated parameters and their estimation uncertainty refer to \ref{app:param}.

%-----------------------------------------------------
%	SIMULATION RESULTS
%------------------------------------------------------
\section{Simulation \label{sec:Simulation}}
\subsection{Hawkes process}
Hawkes processes \cite{HAWKES1971} are a class of multivariate point process which allow for event-occurrence clustering through a stochastic intensity vector. The stochastic intensity is made up from an \textit{exogenous} component where the current intensity is not influenced by prior events and an \textit{endogenous} component where prior events lead to an increased intensity. Within the endogenous component, \textit{self-excitation} refers to an event type leading to more of the same event and \textit{mutual-excitation} is an event driving the occurrence of other event types. 

An $M$-variate Hawkes process is defined as the counting process $N(t) = \{N_m(t)\}_{m=1}^M$ with stochastic intensity $\lambda(t) = \{\lambda^m(t)\}_{m=1}^M$ where
\begin{equation}\label{hawkes:eq:1}
    \lambda^m(t) = \mu^m + \sum_{n=1}^M \int_{-\infty}^t \phi^{mn}(t-s) dN_s^n.
\end{equation}
Here $\mu^m$ is the constant exogenous/background intensity for the $m$th component and $\phi^{mn}(t)$ is the kernel function which governs the dependency of prior events from the $n$th component to the $m$th component at the current time $t$. We use the exponential kernel given as
\begin{equation}\label{hawkes:eq:2}
    \phi^{mn}(t) = \alpha^{mn} e^{-\beta^{mn} t} \mathbbm{1}_{t \in \mathbb{R}^+},
\end{equation}
to encode the dependency on prior events. Here $\alpha^{mn}$ is the scale and $\beta^{mn}$ is the decay for the particular event excitation. This then gives the branching ratios $\Gamma^{mn} = \alpha^{mn}/\beta^{mn}$ which is the average number of events of type $m$ caused by a single event of type $n$.

Simulation is performed using the thinning procedure \cite{TP2012} which can be summarised as follows: let $I^K(t) = \sum_{k=1}^K \lambda^k(t)$ and let $t$ be the current time. We sample an exponential inter-arrival time with rate parameter $I^M(t)$, call it $\tau$. A random uniform $u$ is then sampled from $[0, I^M(t)]$, and if $u < I^M(t) - I^M(t+\tau)$, the arrival time $t + \tau$ is rejected. Otherwise, the arrival time $t + \tau$ is accepted and attributed to the $i$th component, where $i$ is such that $I^{i-1}(t + \tau) < u \leq I^{i}(t + \tau)$.

\subsection{Event types}

The benefit of using Hawkes processes to model the continuous double auction is that each message sent to the exchange by a trader or institution can be thought of as an event. This means that we only need to identify the type of events and specify the corresponding rules for each event.

Orders sent during a simulation are done in a similar manner as that of any other exchange. There are a wide variety of messages that a trader or institution can send to the exchange. These include but are not limited to: Limit Orders (LOs), Market Orders (MOs), cancellations, modifications, Hidden Orders (HOs) and many more. For the purposes of the simulation, we only use the common message types that can be easily identified (without raw message data) and do not result in an excessively complex Hawkes process. The message choices used are: MOs, LOs and cancellations. The LOs and cancellations are further split into aggressive and passive types, leaving us with a total of five message types. Each of these message types are further split into the bid side and ask side, resulting in a total of 10 event types. The various event types used are enumerated in \cref{tab:EventTypes}.

Order modifications are also a commonly used message type. However, including these lead to too many possible event types which results in a Hawkes process that is too complex to reasonably calibrate with a day of data. Therefore, modifications messages are excluded from the simulation.

\begin{table}[htb]
\setlength{\tabcolsep}{4pt}
\centering
\caption{The various event types simulated using the Hawkes process.}
\begin{tabular}{clc} \toprule
Type \# & Event type                            & Side \\ \midrule
1           & Market Order                                                      & Bid \\
2           & Market Order                                                      & Ask \\
3           & \begin{tabular}[c]{@{}c@{}}Aggressive Limit Order\end{tabular} & Bid \\
4           & \begin{tabular}[c]{@{}c@{}}Aggressive Limit Order\end{tabular}  & Ask \\
5           & \begin{tabular}[c]{@{}c@{}}Passive Limit Order\end{tabular}     & Bid \\
6           & \begin{tabular}[c]{@{}c@{}}Passive Limit Order\end{tabular}     & Ask \\
7           & \begin{tabular}[c]{@{}c@{}}Aggressive Cancellation\end{tabular} & Bid \\
8           & \begin{tabular}[c]{@{}c@{}}Aggressive Cancellation\end{tabular} & Ask \\
9           & \begin{tabular}[c]{@{}c@{}}Passive Cancellation\end{tabular}    & Bid \\
10          & \begin{tabular}[c]{@{}c@{}}Passive Cancellation\end{tabular}    & Ask \\ \bottomrule
\end{tabular}
\label{tab:EventTypes}
\end{table}

\subsubsection{Hawkes reference model}\label{sssec:refmod}

Here we do not need any implementation rules to generate order messages because we do not inject messages into the matching engine. Hence there is no need for any implementation mechanics as we can directly calibrate to the simulated data using event types and their associated event times (see section \ref{subsec:Calibration}). The reference model does not produce any volumes or prices associated with the order book events hence cannot be considered a physically meaningful model -- it only generates order event types and their event times. 

\subsection{Implementation rules}

Since a Hawkes process only produces event types and their timings, an associated action must be taken to send a message to the matching engine for each of the event types. We call these the implementation rules. The rules create models that largely drop agent rationality and instead focuses on the problem of understanding the practical implementation constraints. We focus on two separate approaches to managing limit orders via two cases: where limit orders are independent of the spread and where limit orders are dependent on the spread. Both of these cases will share the basic implementation rules for cancellations and market orders.

The basic rules are demonstrated in pseudocode \ref{algo:injectsimulation} in the appendix. These include rules for the implementation of all the order event types in \cref{tab:EventTypes}. We now discuss how these are implemented as actual messages. 

Cancellations require an order identification number (Id.), a side and a price. The order Id is determined by the state of the order book and whether the cancellation is aggressive or passive. The side is determined by the event type. The price is determined by the corresponding order Id. For passive cancellations, we extract all the orders from the appropriate side of the order book depending on the side of the cancellation. A random order with a price that is not equal to the best is sampled and then cancelled. For aggressive cancellations, we extract all the orders from the appropriate side of the order book depending on the side of the cancellation. A random order with a price that is equal to the best is sampled and then cancelled.

Market order messages only require a volume and side. The choice of volume is discussed in \cref{subsec:setup}. The side is determined by the event type. The resulting price is completely determined by the matching engine when it crosses the MO against existing LOs according to the price-time priority strategy. The MO is not sent if there are no orders on the contra side for the MO to cross with. This is to make things easier when cleaning data to identify events in \cref{subsec:identify}. If the MO was sent regardless, the matching engine would simply ignore the MO and we still would not have a transaction. In the case when the MO volume is larger than the combined volume of LOs on the contra side, the MO is only partially filled and the remaining MO is ignored.

Limit order messages require a volume, a side and a price. The choice of volume is discussed in \cref{subsec:setup}. The side is determined by the event type. The price is determined by the state of the order book and whether the LO is aggressive or passive. If the LOB is empty from the same side as the initiating LO, the price is then set to be a random integer from 1 to 10 ticks away from the best on the contra side. For example, if we have a LO from the bid side (either event type 3 or 5) and the bid side of the LOB is empty, we set the price to be 1 to 10 ticks below the best ask. In the case when both sides of the LOB are empty, we apply the same rule as above, with the exception that the reference price on the contra side uses the previous best bid or ask before the LOB became empty.

For passive LOs, we set the price to be one tick worse than the current best. For example, if we have a passive LO from the bid side (event type 5), we set the price to be one tick below the best bid. 
For aggressive LOs, we have two rules to set the price which results in two separate models.
This implies that the Hawkes model based simulation here will lead to three distinct models: i.) the Hawkes reference model in section \ref{sssec:refmod} for the events and their times, ii.) Model 1 and iii.) Model 2 for the different implementations of aggressive limit orders.

\subsubsection{Limit order model 1} \label{sssec:model 1}

{\it This model has no mechanism controlling for the spread}. Here the price is always set to be one tick better than the current best even if it results in the new best crossing the contra side. For example, if we have an aggressive LO from the bid side (event type 3), we set the price to be one tick above the best. The rules are demonstrated in pseudocode \ref{algo:model1}.

For this model, the matching engine handles the crossing of LOs by converting the crossing LO into a MO to execute as a trade. The excess volume that did not execute as a trade remains in the LOB as a limit order at its original crossing price. For example, consider the best ask being a single order at a price of 50 with a volume of 30. Suppose a new LO bid is placed at a price of 50 and a volume of 70. Then a volume of 30 from the new LO will be executed as a trade against the best ask resulting in a trade at a price of 50 with a volume of 30. The remaining volume of 40 from the LO bid now becomes the best bid with a price of 50.

\subsubsection{Limit order model 2} \label{sssec:model 2}

{\it The limit price for this model is dependent on the current spread.} If the spread is greater than one tick, the price is set to be one tick better than the current best. Otherwise if the spread is equal to one tick, the price is set to be the same as the current best. This model controls the spread to ensure that LOs do not cross the order book. For example, if we have an aggressive LO from the bid side (event type 3), we set the price to be one tick above the best if the current spread is greater than one tick. Otherwise we set the price to be the same as the current best if the current spread is equal to one tick. The rules are demonstrated in pseudocode \ref{algo:model2}.

\subsection{Simulation Initialisation}\label{subsec:setup}

The simulations implemented are restricted to the continuous trading session with order types only being limit, market and cancel orders which have a DAY time-in-force.\footnote{The DAY TIF is the default for all orders and implies that these orders will only be valid for the day they are submitted.} A single client and stock is adopted for simplicity to represent the market as a whole. The simulation is performed in Julia by using the simulated 10-variate Hawkes and submitting the appropriate order to the matching engine at the correct time based on the event type and the rules defined. The simulation rules require that the client receive market data feedback from the matching engine in the form of a snapshot of the limit order book and the current best bid/ask. 

Timeouts may occur if the inter-arrival times are less than the time it takes for the matching engine to process the data plus the time it takes for Julia to receive market data and process the next order. This occurs if there was a large build up of orders which affects processing time and round-trip latency. Cancellation orders are the most affected because this requires running through all the orders on one side of the LOB to determine if the price of an order is equal or not equal to the best, to then find the appropriate sample of orders from which we can cancel. It is therefore important that we have appropriate Hawkes parameters and volume choices to ensure that neither model has a large build up of orders.

Having a matching engine at our disposal means that simulations need not be conducted in real-time, {\it i.e.,} we can speed up the simulation time relative to real business or calendar time assuming that we do not force timeouts. Furthermore, data output can be modified when needed which can be convenient for debugging and analysis.

\subsubsection{Parameter choice}

The parameters for the Hawkes process used in the simulation were not calibrated to real data. This was a pragmatic choice. When calibrating with real data we need to worry whether we have the correct specification and then whether we have an indicative window of data with enough events to empirically capture the specification under estimation. Finding the right specification may inadvertently lead to a Hawkes process that is more complicated than necessary which may lead to estimation complications later on -- for example using a sum of exponential kernels significantly increases the number of parameters required. This would not change the arguments made in this work -- it would just add unnecessary complexity. 

In tandem with model complexity is the issue of having sufficient observations of all event types of interest. Many datasets may have periods of insufficient order types to calibrate the process for particular scenarios that may have features of interest. For example, the A2X exchange has many days where there are less than 10 transactions per day. Therefore, it has insufficient market orders for calibration. Excluding market orders is not always an option because it is can be an essential feature of the model.

The problem is generic in that Hawkes process modelling seems effective with regards to creating a model snap-shot of a particular set of market conditions in a particular window of data but requires sufficiently representative data for identification. It can be a powerful visual diagnostic tool that can relate order types under a particular set of market conditions. However it becomes difficult to create simple models that capture the dynamics of interest with a sparse finite piece of data.

Another reason why we did not calibrate to real data is to ensure that we achieve a correct balance in liquidity. Incorrectly balancing liquidity can cause latency problems when receiving market data from the matching engine which can cause timeouts in our simulation.\footnote{This is a data monitoring restriction rather than an order book latency problem and required an enhancement of the requests associated with LOB snapshots. It is recommended that the LOB does not get too big, {\it i.e.,} $>$ 100 orders on each side.} For this reason extracting data snapshots of the LOB under simulation was a key area requiring some additional development work.\footnote{A problem was also identified in the processing of LOs that crossed the spread, whereby the contra side would be executed against the crossed LO instead of vice versa. This had no affect on the matching dynamics but impacts the trade reporting.} \footnote{The reporting of trades and order cancels was improved by assigning their IDs that of the corresponding order that was executed.}

The parameters we used were chosen to ensure that we maintain balance between events that provide liquidity (event types 3--6) and events that remove liquidity (event types 1--2 and 7--10). Obtaining this balance is important because if there are too many events that take away liquidity then the LOB will often be empty. If there are too many events that provide liquidity then there will be a large build up of orders, particularly for model 2, which can result in no movements in the price. The Hawkes process we implemented was then with a single exponential kernel with parameters given as: $\boldsymbol{\alpha} = \boldsymbol{\mu} \boldsymbol{1}^\top$, $\boldsymbol{\beta} = \{\beta^{mn}\}_{m=1,n=1}^{10}$ where $\beta^{mn} = 0.2$ for all $m$ and $n$, and the baseline intensity $\boldsymbol{\mu}$ as the vector :
\begin{equation}
    [0.01, 0.01, 0.02, 0.02, 0.02, 0.02, 0.015, 0.015, 0.015, 0.015]^\top.\nonumber
\end{equation}
The challenge of balancing liquidity is perhaps a key motivation for moving away from point-process type models to models that can capture a more strategic and dynamic interplay between market participants.

\begin{figure*}[htb]
    \centering
    \subfloat[Model 1 - Full 8 hour day]{\includegraphics[width=.49\textwidth]{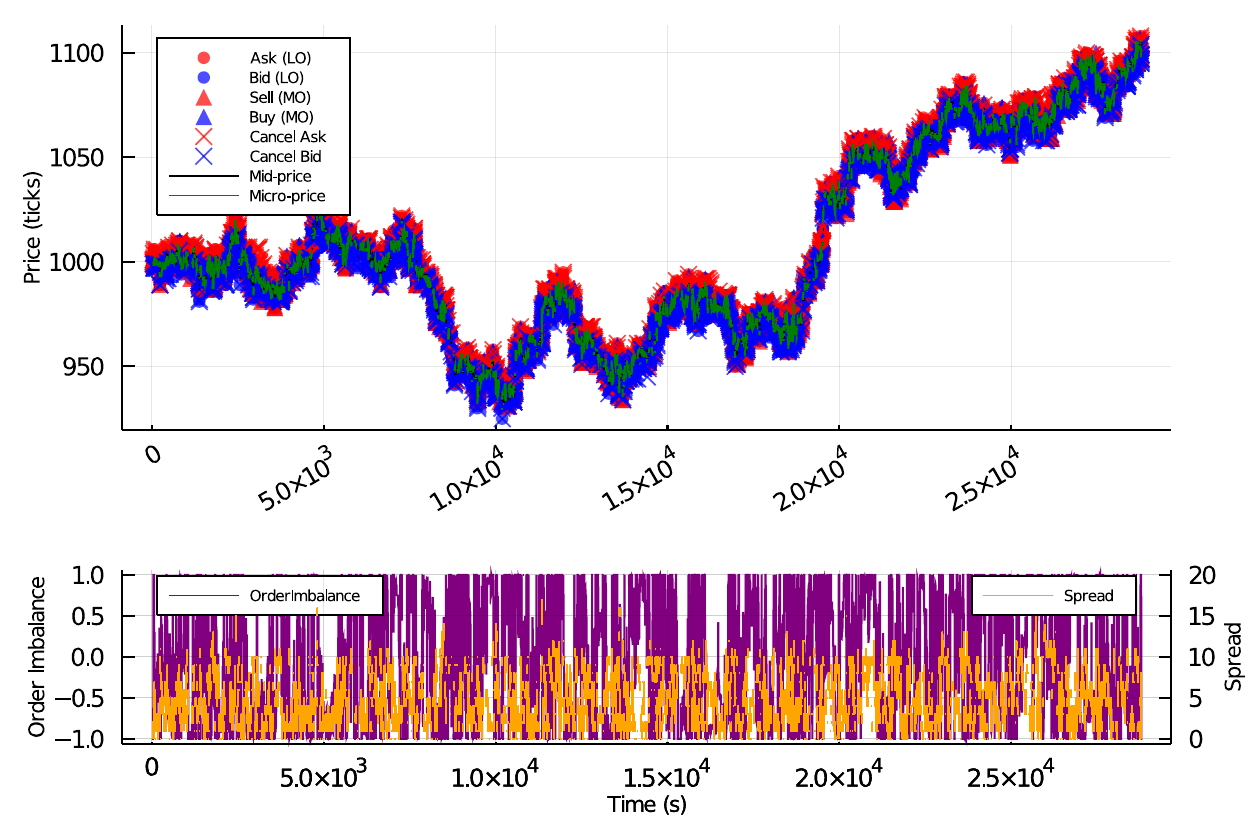}}
    \subfloat[Model 1 - 1 hour]{\includegraphics[width=.49\textwidth]{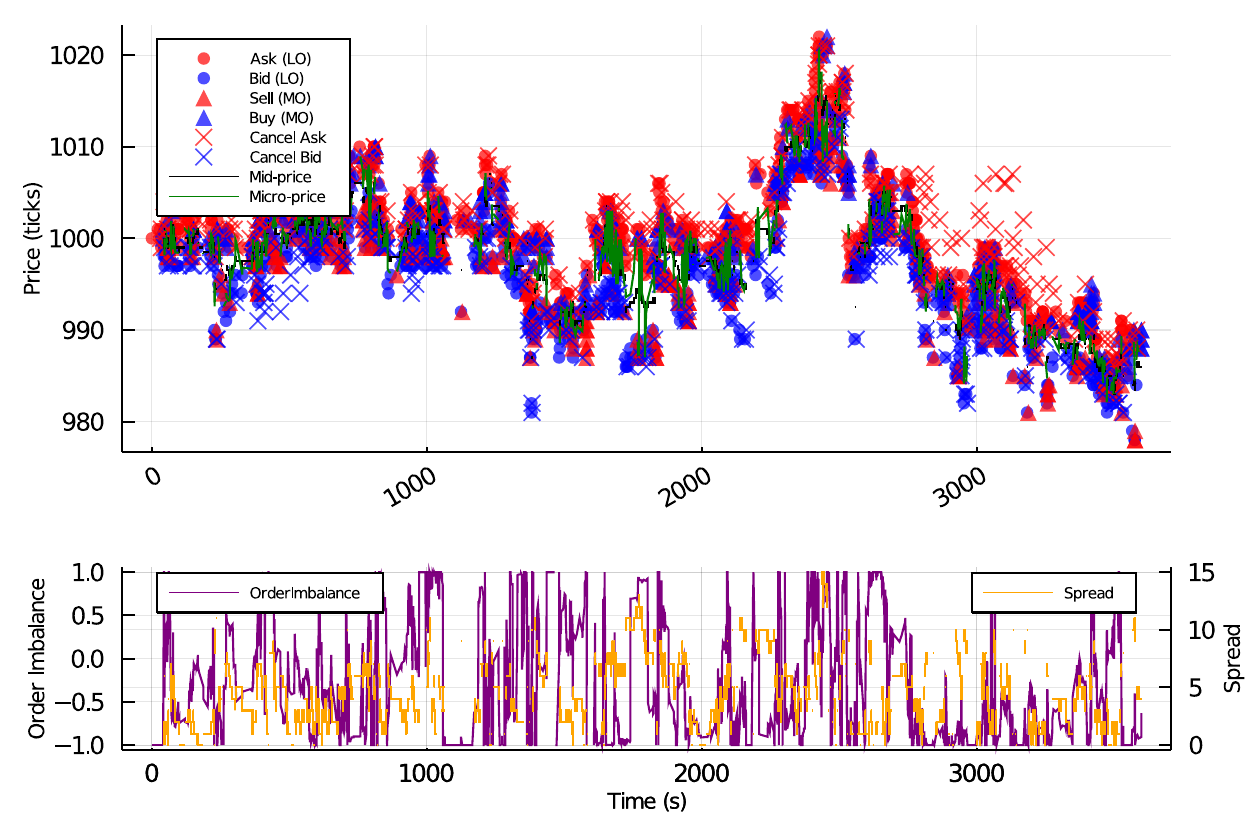}} \\
    \subfloat[Model 2 - Full 8 hour day]{\includegraphics[width=.49\textwidth]{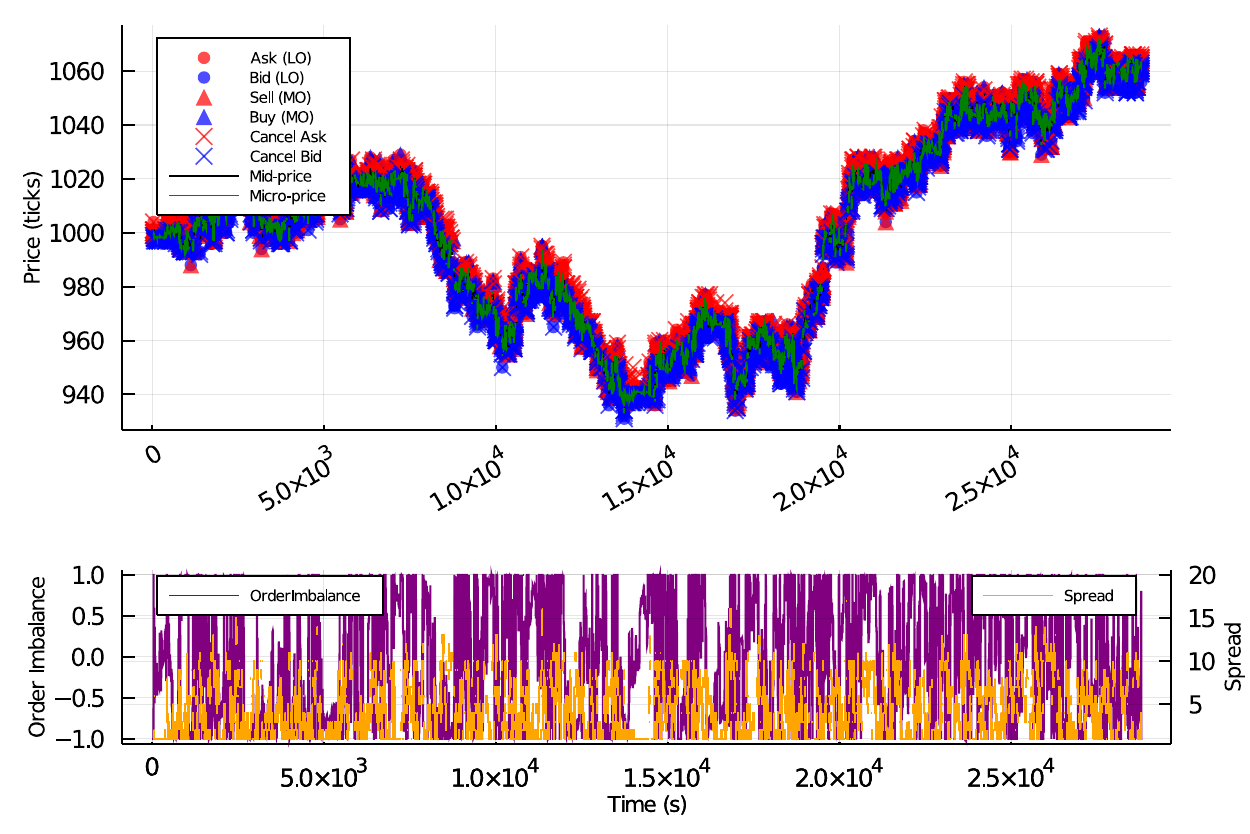}}
    \subfloat[Model 2 - 1 hour]{\includegraphics[width=.49\textwidth]{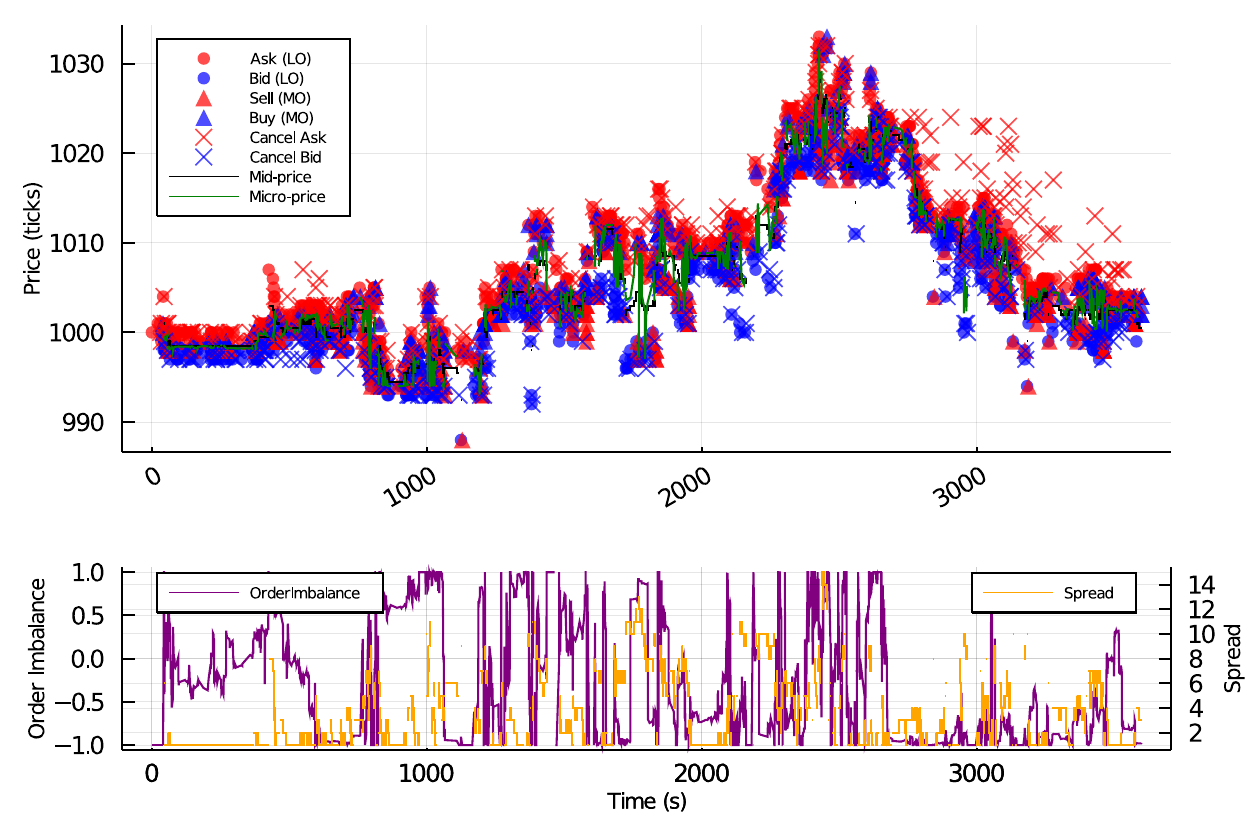}}
    \caption{The simulation of a 10 variate Hawkes process used to generate the different order types submitted to CoinTossX in table \ref{tab:EventTypes} is visualised over an 8-hour and 1-hour period and compared across both models. Together with the plot of orders submitted (top), we visualise the evolution of the spread (bottom, right axis) and the order imbalance (bottom, left axis).}
    \label{fig:sim}
\end{figure*}

\subsubsection{Volume choice}

The final component required to conduct simulations is a choice for the volumes associated with MOs and LOs. There seems to be no consensus for a model specification for the volume of orders as the distributions varies widely with the product and market. However, it was empirically found that the unconditional distribution of orders follows a power-law behaviour \cite{CTPA2011a}.

Therefore, we will assume that the volume associated with each LO and MO is from an independent and identically distributed (IID) power-law where the probability density function is given as:
\begin{equation}\label{eq:14}
    f(x)=
    \begin{cases}
    \frac{\alpha x_{\mathrm{m}}^{\alpha}}{x^{\alpha+1}} & x \geq x_{\mathrm{m}} \\
    0 & x<x_{\mathrm{m}}.
    \end{cases}
\end{equation}
We set $x_{\mathrm{m}}$ to be 20 for LOs and 50 for MOs, whereas we set $\alpha = 1$ for both MOs and LOs. These choices along with our parameter choices ensured that there was never a large build up of orders in model 2.

The volume sizes play an important role in ensuring that there is not a large build up of orders for model 2.\footnote{Model 1 does not suffer from a large build up of orders because the model permits LOs to cross the spread. Therefore, this model has a natural way to remove too much liquidity.} In particular, the total MO volume cannot be too small relative to the total LO volume. This is because even though there are cancellations to assist in removing liquidity, cancellations can also occur on partially filled LOs which can then still result in a build up of orders. Moreover, depending on the state of the order book, MOs and cancellations are sometimes lost because there is nothing to execute against or nothing to cancel, whereas LOs will always be placed.

\subsection{Simulation results}
Simulations for both model 1 and model 2 are performed using the same realisation of the Hawkes reference model with the same event types, counts and volumes. The simulations were performed using pseudocode \ref{algo:injectsimulation} and the only difference between the two models are in the rules for limit order price placement.

\Cref{fig:sim} plots the resulting simulation of the two models. The first and second rows are model 1 and 2 respectively, and the first and second columns are the full 8 hour day and the first hour respectively. Additionally, the evolution of the mid-price, micro-price, spread and order-imbalance is tracked to gain a better view of the underlying dynamics. The following dynamics can be thought of being representative of a low-liquidity market, given the event counts in table \ref{tab:EventCounts}.\footnote{The order imbalance is given by \cite{cartea2015algorithmic} $\rho_t = \frac{v^b_t - v^a_t}{v^b_t + v^a_t}$ where $v^b_t$ and $v^a_t$ are the total bid and ask side volume of the LOB at time $t$.}

Despite there being significantly more market orders for model 1 (refer to table \ref{tab:EventCounts}), both models' price series were found to be fairly similar. Despite this similarity, model 1 exhibits a slightly greater price range by the end of the day. The spread remained erratic throughout the simulation period, reaching as high as 15 ticks. Over the shorter time-period, as expected, breaks on either side of the LOB would result in more aggressive price movements than would otherwise occur in a high liquidity market.

%-----------------------------------------------------
%	ESTIMATION RESULTS
%-----------------------------------------------------
\section{Estimation \label{sec:Estimation}}

\subsection{Event identification}\label{subsec:identify}
In order to compare and investigate how market mechanisms may affect our ability to infer the underlying process we first need to classify the resulting data from the models appropriately. We use the same set of event types in \cref{tab:EventTypes}.

When classifying market orders we need to be careful. Trades with a large volume submitted will walk the LOB. These will execute at multiple price levels and are split into multiple trades corresponding to the the different orders against which they are executed in the reported data. These transactions will also all report the same timestamp. Therefore, these transactions must be aggregated and counted as one event.

Aggressive limit orders are any of those (a) whose limit price is equal to the current best, (b) whose limit price is better than the best, or (c) that occur in an empty LOB -- thereby becoming the new best. Passive limit orders are those whose limit price is worse than the current best.

Cancel orders are classified as aggressive if the order cancelled was in the L1LOB (at the best). Otherwise the cancellation is classified as passive.
%%%%%%%%%%%%%%%%%%%%%%%%%
% Market Mechanics
\subsection{Distortions due to market mechanics}\label{ssec:mech}

From practical market mechanic considerations dependant on the state of the order book we can explain why certain classified events arising from the matching engine deviate from the true event type injected into the matching engine. There are several cases that contribute towards this distortion.

First, from pseudocode \ref{algo:model1} and \ref{algo:model2}, both passive and aggressive LOs can be submitted to an empty LOB.\footnote{This typically only occurs in markets with very little liquidity ({\it e.g.} A2X).} This order immediately becomes the best which means that passive limit orders can become aggressive.

Second, cancelled orders will not go through the matching engine and be reflected in the data feed when there are no appropriate orders to cancel. In particular, both passive and aggressive cancellations will not be executed when the LOB is empty on the side which the cancellation came from. Alternatively, passive cancellations cannot cancel orders when there is only price level (the best) in the LOB. Similarly, market orders are ignored when it arrives while the contra side of the LOB is empty.

Lastly, model 1 requires additional care for limit orders that cross the spread. Limit orders that cross the spread become an effective market order. Additionally, the crossing can spawn an additional event which results in an effective market order and an aggressive limit order. This only occurs when the crossed limit order was only partially executed against the contra side, \textit{i.e.,} when the incoming LO has a volume larger than the best on the contra side. 

The case when an additional event is spawned becomes problematic because these two events will have the same timestamp. Therefore, to overcome the problem of concurrent events in a point process we add one millisecond to the additional aggressive limit order that was spawned. We can do this because we know that for the matching engine, limit orders that cross will first execute as a market order before it becomes a limit order with a reduced volume. Since we know the true ordering of events, we can make the small adjustment to ensure that the data remains a simple point process (see \citet{L2007} and the discussion around concurrent events).

\begin{table}[htb]
\setlength{\tabcolsep}{4pt}
\centering
\caption{The various event counts for the three models.}
\begin{tabular}{cccc} \toprule
\shortstack{Type\\number} & \shortstack{Hawkes\\count} & \shortstack{Model 1\\count} & \shortstack{Model 2\\count} \\ \midrule
1           & 1436 & 1533 & 1228 \\
2           & 1380 & 1579 & 1268 \\
3           & 2873 & 3071 & 3164 \\
4           & 2875 & 3046 & 3152 \\
5           & 2736 & 2356 & 2445 \\
6           & 2726 & 2360 & 2448 \\
7           & 2135 & 1841 & 1918 \\
8           & 2075 & 1799 & 1866 \\
9           & 2095 & 1379 & 1490 \\
10          & 2068 & 1393 & 1517 \\ \bottomrule
\end{tabular}
\label{tab:EventCounts}
\end{table}

\Cref{tab:EventCounts} reports the event counts from the Hawkes reference model and the events classified from model 1 and 2. We see that the event counts from both models deviate from the Hawkes reference model used to create the models. The key question is: Do these market mechanisms and practical considerations affect our ability to infer the underlying process generating the models?

%%%%%%%%%%%%%%%%%%%%%%%%%%%
% CALIBRATION
\subsection{Calibration \label{subsec:Calibration}}
We use the popular parametric approach to calibrate a Hawkes process using the Maximum Likelihood Estimation (MLE) \cite{BMM2015,TP2012}. To calibrate we maximise the log-likelihood (the log-likelihood and the optimisation details are given in \ref{app:likelihood}):
\begin{equation}%\label{hawkes:eq:10}
    \hat{\theta} = \underset{\theta}{\text{argmax }} \ln \mathcal{L}\left( \theta \right), 
\end{equation}
where $\theta$ refers to the parameters of the Hawkes process which include $(\mu^m, \alpha^{mn}, \beta^{mn})$, depending on the specification of the model. Here we do not enforce a specific design on the matrices $\boldsymbol{\alpha}$ or $\boldsymbol{\beta}$ and allow the calibration to determine the structure of the matrices. Based on the events identified from table \ref{tab:EventCounts} we have 10 event types and therefore we have a total of $10 \times 10 \times 2 + 10 = 210$ parameters. 

We perform the calibration using MLE to find $\hat{\theta}_{\text{Hawkes}}$, $\hat{\theta}_{\text{M1}}$ and $\hat{\theta}_{\text{M2}}$ for the Hawkes reference model, and for model 1 and 2, respectively, with initial parameter values set to the true starting values $\theta_{true}$.

The results are summarised by measuring the deviation between the MLE parameters $\hat{\theta}$ against the original Hawkes parameters $\theta_{\text{true}}$ from \cref{subsec:setup} using the mean absolute error (MAE)\footnote{$\text{MAE}(\hat{\theta},\theta_{\text{true}}) = \frac{1}{210} \sum_{k = 1}^{210} |\theta_{\text{true},k} - \hat{\theta}_k|.
$}
and the root mean square error (RMSE)\footnote{$\text{RMSE} = \sqrt{\text{MSE}}$ and $\text{MSE}(\hat{\theta},\theta_{\text{true}}) =  \frac{1}{210}\sum_{k = 1}^{210} \left( \theta_{\text{true},k} - \hat{\theta}_k \right)^2.
$}.

A detailed visualisation of the parameters can be found in \ref{app:param}.

\begin{table}[htb]
    \caption{Measures of deviation between Hawkes input parameters and calibrated parameters for the Hawkes reference model as well as for model 1 and 2 simulations.}
    \centering
    \begin{tabular}{cccc} \toprule
         Measure & Hawkes & Model 1 & Model 2 \\ \midrule
         MAE  & 0.0542 & 16.3300  & 25.8408  \\
         RMSE & 0.1439 & 107.0479 & 224.6096 \\ \bottomrule
    \end{tabular}
    \label{tab:distortion}
\end{table}

\Cref{tab:distortion} reports the MAE and RMSE of the calibrated parameters for the Hawkes reference model and for model 1 and 2 against the original Hawkes parameters. We see that the calibrated parameters from the reference model present the smallest deviation from the original parameters. This is expected because there are no market mechanisms or practical considerations changing the event types, therefore the small deviation is an indication that our calibration procedure is correct. 

\begin{figure*}[htb]
    \centering
    \subfloat[Hawkes]{\includegraphics[width=.33\textwidth]{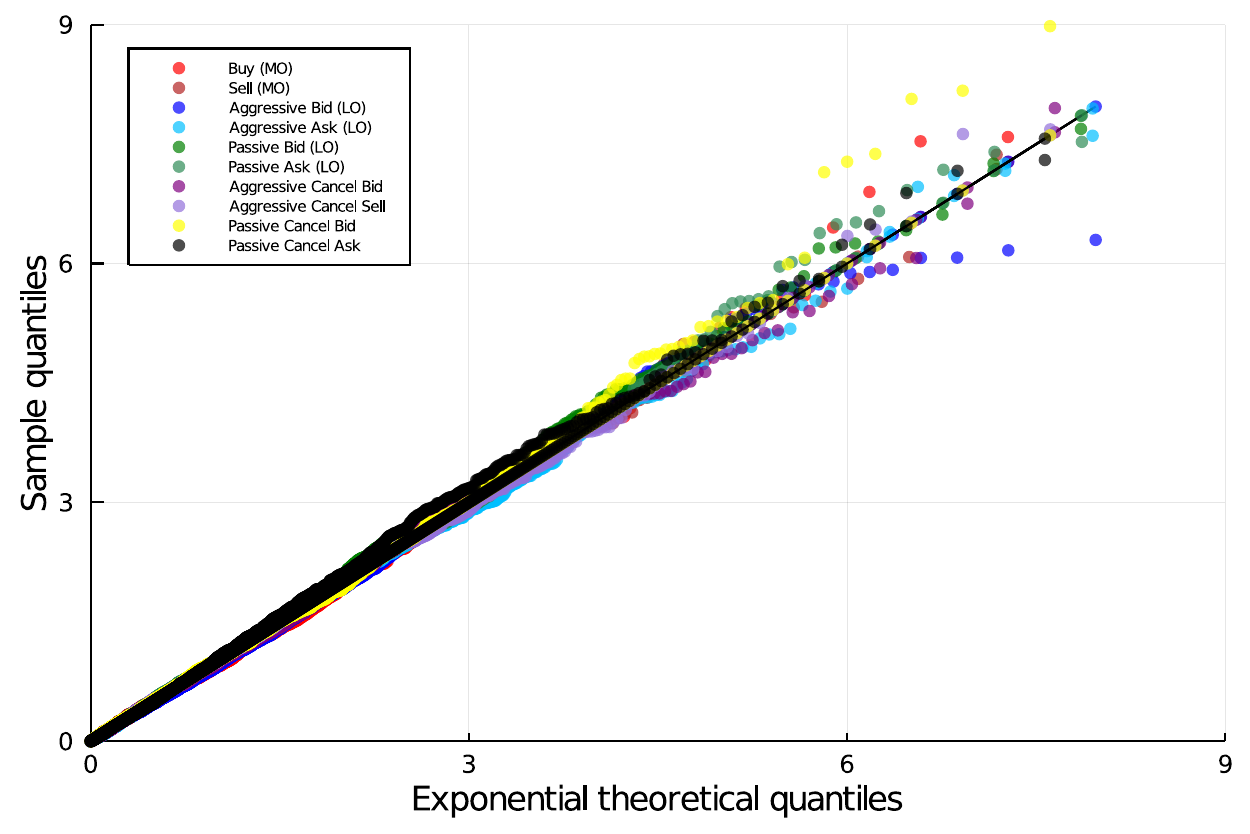}}
    \subfloat[Model 1]{\includegraphics[width=.33\textwidth]{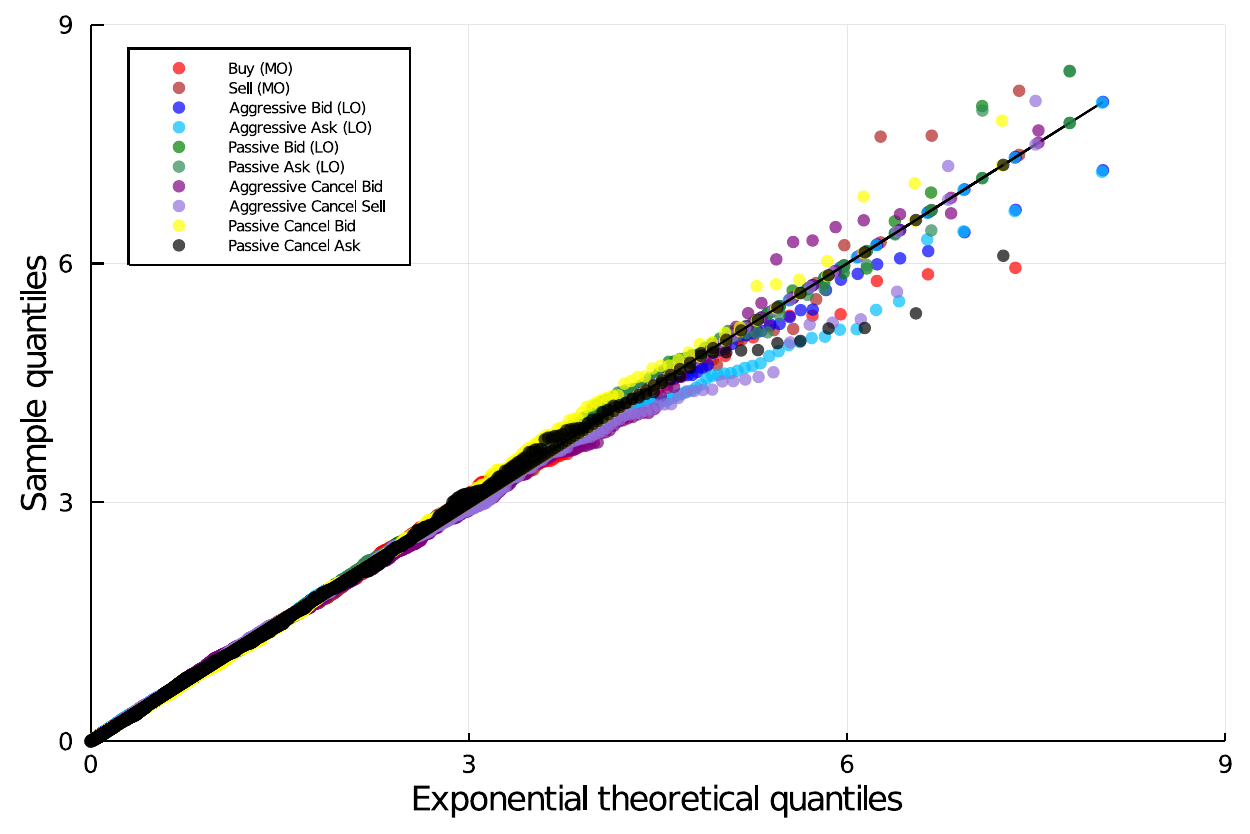}}
    \subfloat[Model 2]{\includegraphics[width=.33\textwidth]{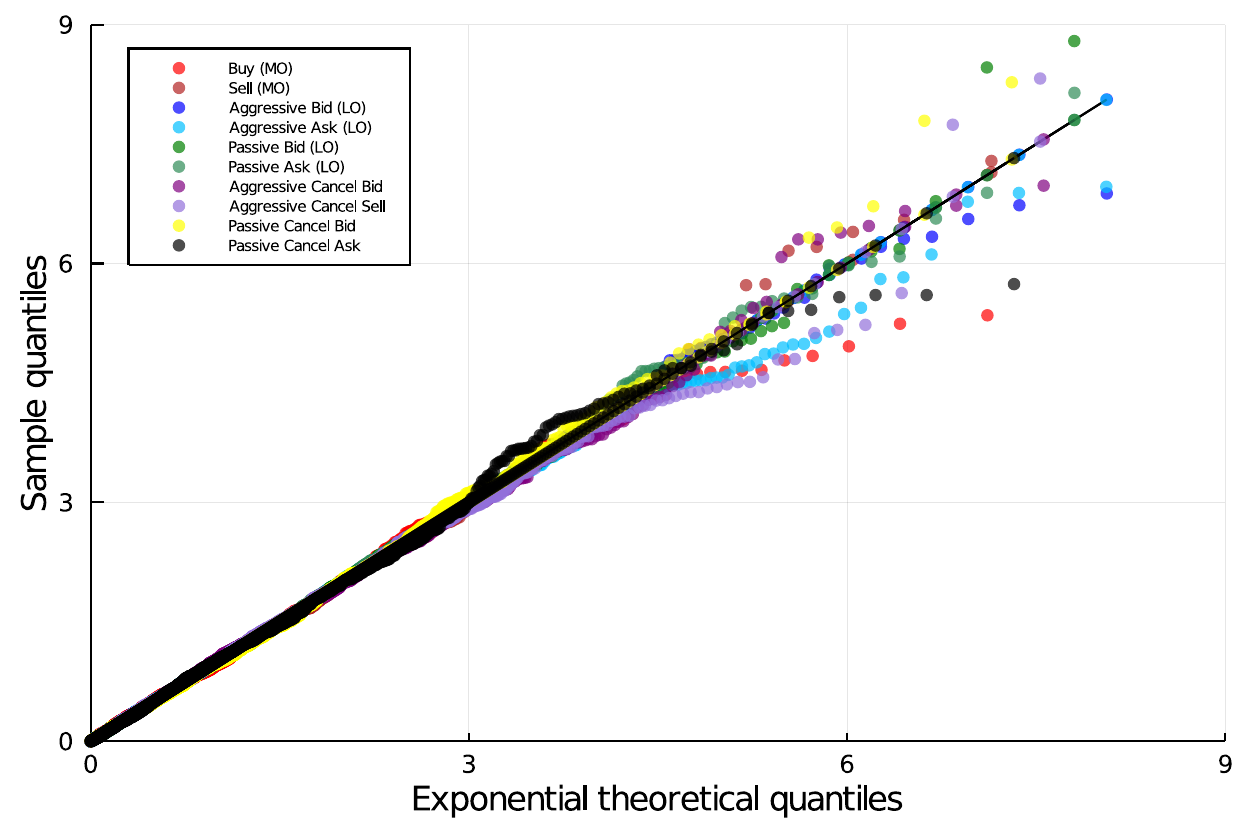}} \\
    \caption{As a qualitative measure of goodness-of-fit, generalised residuals for each event type (obtained from \cref{hawkes:eq:11}) are compared against the quantiles of an Exponential distribution with unit rate.}
    \label{fig:QQ}
\end{figure*}

\begin{table*}[htb]
\setlength{\tabcolsep}{4pt}
\caption{Mis-specification tests are performed by checking the null hypotheses that generalised residuals are serially independent and are exponentially distributed with unit rate. The values reported are $p-$values.}
\centering
\begin{tabular}{ccccccc} \toprule
{} & \multicolumn{2}{c}{Hawkes} & \multicolumn{2}{c}{Model 1} & \multicolumn{2}{c}{Model 2} \\ \cmidrule(lr){2-3} \cmidrule(lr){4-5} \cmidrule(lr){6-7}
\shortstack{Type\\number} & \shortstack{Ljung\\-Box} & \shortstack{Kolmogorov\\-Smirnov} & \shortstack{Ljung\\-Box} & \shortstack{Kolmogorov\\-Smirnov} & \shortstack{Ljung\\-Box} & \shortstack{Kolmogorov\\-Smirnov} \\ \midrule
1  & 0.116   & 0.5279  & 0.32051 & 0.87332 & 0.96014 & 0.5663 \\ 
2  & 0.455   & 0.23593 & 0.57889 & 0.69011 & 0.98565 & 0.90626 \\
3  & 0.51286 & 0.13925 & 0.86472 & 0.75525 & 0.41731 & 0.97167 \\
4  & 0.14084 & 0.63515 & 0.51574 & 0.15786 & 0.85563 & 0.72569 \\
5  & 0.54962 & 0.55145 & 0.73988 & 0.89928 & 0.85563 & 0.94349 \\
6  & 0.64359 & 0.20044 & 0.3429  & 0.83269 & 0.58501 & 0.64426 \\
7  & 0.16513 & 0.75415 & 0.81173 & 0.34173 & 0.00777 & 0.57761 \\
8  & 0.92577 & 0.39957 & 0.49469 & 0.53315 & 0.59695 & 0.67609 \\
9  & 0.1144  & 0.88623 & 0.5002  & 0.60972 & 0.90512 & 0.65956 \\
10 & 0.56949 & 0.73368 & 0.33559 & 0.69256 & 0.39552 & 0.73026 \\ \bottomrule
\end{tabular}
\label{tab:HypothesisTests}
\end{table*}

The calibrated parameters from model 1 and 2 both deviate significantly from the original parameters. This is expected because from \cref{tab:EventCounts} we see that model 1 and 2 are affected by market mechanisms and practical considerations which cause additional events to spawn, events being dropped and event types changing. Therefore, the large deviation is an indication that market mechanisms and practical considerations have changed the process such that it is no longer the same as the original process.

We will corroborate this using a hypothesis test in \cref{sec:discussion}. First we will ensure that our calibrated parameters have been correctly identified by using the multivariate random time change theorem proposed by \citet{BOWSHER2007}.

\subsection{Validation}\label{ssec:validation}

To check the validity of our calibration we utilise the multivariate random time change theorem \cite{BOWSHER2007}.

\begin{theorem}\label{hawkes:MRTC}
Theorem 4.1 of \citet{BOWSHER2007} (rephrased):  \\
Consider the $M$ sequences of generalised residuals $\{e_i^{(m)}(\theta)\}_{m=1}^M$, where
\begin{equation}\label{hawkes:eq:11}
    e_i^{(m)}(\theta) = \int_{t_i^m}^{t_{i+1}^m} \lambda^m(s;\theta) ds.
\end{equation}
When $\theta$ is the true set of parameters, then each sequence $e_i^{(m)}(\theta)$ will be an independently distributed exponentially random variable with unit mean.
\end{theorem}

Using this result by computing \cref{hawkes:eq:11} with $\hat{\theta}$ from the calibration, we can perform statistical tests to determine whether the generalised residuals are in fact independent and exponentially distributed with unit rate. To this end, we first perform a simple Q-Q (quantile-quantile) plot for visual inspection. 

We then perform the Kolmogorov-Smirnov test which is a non-parametric test based on the maximal discrepancy between the empirical cumulative distribution and the exponential cumulative distribution with unit rate. An \textit{acceptance} of the null hypothesis $H_0$ for this test indicates that generalised residuals are exponentially distributed with unit rate.

Finally, the Ljung-Box test can be used to test the independence. The test examines the null hypothesis of an absence of auto-correlation in a given time-series. For significance level $\alpha$, the critical region for rejection of the hypothesis of randomness is
\begin{equation}
Q = n (n + 2) \sum_{k = 1}^{h} \frac{\hat{\rho}^2_k}{n - k} > \chi^2_{1 - \alpha, h}
\end{equation}
where $n$ is the sample size, $\hat{\rho}_k$ is the sample auto-correlation at lag $k$, and $Q$ asymptotically follows a $\chi^2_{h}$ distribution. An \textit{acceptance} of the null hypothesis $H_0$ for this test indicates that generalised residuals are serially independent.

\Cref{fig:QQ} plots the Q-Q plot of the generalised residuals for each of the event types against the theoretical quantiles from an exponential with unit rate. From left to right, we have the generalised residuals from the reference model, model 1 and model 2 respectively. From a visual perspective, we see that the generalised residuals for all three calibrations seem to follow the quantiles from the theoretical exponential.

For more quantitative results, \cref{tab:HypothesisTests} reports the $p-$values from the Ljung-Box and the Kolmogorov-Smirnov test for the generalised residuals from each event type for each calibration. We see that all of the $p-$values (except the Ljung-Box test on event type 7 from model 2) are larger than a 10\% threshold. Therefore, the various null hypotheses $H_0$ are accepted which means that our generalised residuals are indeed independent and exponentially distributed with unit mean.

The successful application of \cref{hawkes:MRTC} is an indication that the results obtained from the MLE are indeed true parameters for their respective data. Coupling this result along with the result from \cref{tab:distortion}, it indicates that market mechanisms have indeed altered the underlying process such that it is no longer the same as the initial Hawkes process which we used to generate the various event types.

%-----------------------------------------------------
%   RESULTS
%-----------------------------------------------------
\section{Discussion of results}\label{sec:discussion}

\subsection{Hypothesis tests}\label{ssec:hyptest}
In order to statistically determine if market mechanisms have altered the underlying process, we can perform a simple vs composite hypothesis test using the likelihood-ratio test also known as Wilks test \cite{Wilks1938}.

We can do this by testing
\begin{equation}\label{eq:12}
    H_0: \theta = \theta_{\text{true}} \quad \text{vs} \quad H_1: \theta \in \Theta,
\end{equation}
where $\Theta$ is the full unrestricted parameter space. Therefore, the test statistic for the hypothesis test is given by
$$
\lambda_{\text{LR}} = -2 \ln \left[ \frac{\mathcal{L}\left( \theta_{\text{true}} \right)}{\sup\limits_{\theta \in \Theta} \mathcal{L}\left( \theta \right)} \right],
$$
which reduces to
\begin{equation}\label{eq:13}
    \lambda_{\text{LR}} = -2 \left[ \ln \mathcal{L}\left( \theta_{\text{true}} \right) - \ln \mathcal{L}\left( \hat{\theta} \right) \right],
\end{equation}
where $\hat{\theta}$ are the estimates obtained through MLE. The test statistic in \cref{eq:13}, subject to regularity, will be approximately distributed as $\chi^2_{210}$ assuming the null hypothesis is true \cite[p.~112--114]{Silvey1975}.

\begin{table}[htb]
    \caption{Test statistic and $p-$values for the likelihood-ratio test of \cref{eq:12}.}
    \centering
    \begin{tabular}{ccc} \toprule
         Model      & $\lambda_{\text{LR}}$ & $p-$value  \\ \midrule
         Hawkes     & 235.50                & 0.109457   \\
         Model 1    & 4636.38               & 0          \\
         Model 2    & 1197.95               & 0          \\ \bottomrule
    \end{tabular}
    \label{tab:LRtest}
\end{table}

\Cref{tab:LRtest} reports the test statistic and $p-$values for the likelihood-ratio test on the reference model and on model 1 and 2. The results from the reference model indicate that it remains probable that the observed data came from $\theta_{\text{true}}$. However, the results from model 1 and 2 indicate that is essentially impossible that the observed data came from $\theta_{\text{true}}$.

This is expected since the reference model is simply a realisation of $\theta_{\text{true}}$, whereas model 1 and 2 are affected by market mechanisms that drop events, change event types and spawn additional events. These results confirm that market design and its mechanisms can indeed affect our ability to infer the underlying process. 

\subsection{Implications}

% We have demonstrated that market mechanics and practical constraints can significantly cloud our ability to identify the true process used to generate orders because the process in the given representation does not fit the interplay between orders and the matching engine. Even when the true process is provided it is change in a statistically meaningful way by the matching process -- and will always do so. The changes to the resulting orders and trades due to the market mechanics were significant enough, according to a likelihood-ratio test, to make it is practically impossible to conclude the resulting data came from the same process that was used to generated the data. It can be argued that the clouding of the order generating process by a matching engine is a triviality. However, this is not the case. We argue that it suggests that the data generating process may not be faithfully represented using the model's framework itself. %However, we note that not matter what model is chosen it is unlikely to remain identifiable for the two cases provided.

We have demonstrated that market mechanics and practical constraints can significantly cloud our ability to identify the true process used to generate orders when the given representation does not fit the interplay between orders and the matching engine. When this is the case, the process is changed in a statistically meaningful way by the matching process. The changes to the resulting orders and trades due to the market mechanics were significant enough, according to a likelihood-ratio test, to make it is practically impossible to conclude that the resulting data came from the same process used to generate the data. It may be argued that the clouding of the order generating process by a matching engine is trivial. However, this is not the case. We argue that it suggests that the data generating process may not always be faithfully represented using the model's framework itself.

These results suggest that care should be taken when applying alternative market models, such as agent-based models, to continuous double auctions. Specifically, that agent behaviour can also be dominated by practical considerations rather than strategic interactions alone. Therefore, models need to be able to account for this and cannot be independent of the market rules and mechanisms.

For the example that we used, this can possibly be ameliorated by using a constrained Hawkes process similar to what was done by \citet{zheng2014modelling}. We can also introduce new processes to keep track of the spread and the number of orders on each side of the LOB. These processes can inform us which intensities to turn off so as to ensure that event types which would otherwise violate practical considerations and market rules do not occur. Ideally, these types of additional complexities should arise naturally from a model specification rather than added as constraints.

The considerations that one would need to account for will depend on the model and the set of actions that can be taken. Moreover, the considerations will also depend on the specific market and their specific rules. For example, in 2003, Chinese stock markets only allowed limit orders to be submitted and there were no ``true'' market orders \cite{Zhou2012}. Therefore, the considerations will need to be tailored for the model and its application.

This does not detract from the value that a point-process model has in testing trading infrastructure as it can provide fast exploratory data analytics. The work here successfully demonstrates this use with the CoinTossX matching engine on test cases that have reasonable realism in terms of the volume and intensity of the order flow with different order types.

% This does not detract from the value that such a point-process model has in testing trading infrastructure to provide fast exploratory data analytics to better understand the dynamics of the order book. In part, the work here successfully demonstrates the speed and dynamics of the CoinTossX matching engine on test cases that have reasonable realism in terms of the volume and intensity of the order flow with different order types.

%------------------------------------------------------
%	CONCLUSION
%------------------------------------------------------
\section{Conclusion \label{sec:Conclusion}}

We investigate whether market microstructure and the mechanics of the order book can cloud our ability to identify the underlying process when it is known, and demonstrate the ability of the CoinTossX matching engine to manage realistic orders in an expected manner. This work is an important milestone to test the matching engine, and the initial development of the model management system that will be used to generate agent-based models predicated on a continuous double auction in future work. Here we use a 10-variate Hawkes process along with simple rules to generate order messages that are injected into a matching engine for order matching and limit order book aggregation and management.

From the simulation we compare the calibrated parameters for a Hawkes reference model (section \ref{sssec:refmod}) with two additional models (sections \ref{sssec:model 1} and \ref{sssec:model 2}) with modified implementation rules against the true model parameters (see table \ref{tab:distortion} in section \ref{subsec:Calibration}). We find that the calibrated parameters from the models deviate significantly from the true parameters (see section \ref{subsec:Calibration}). The calibration is then validated using the multivariate random time change theorem to ensure that the calibrated parameters are correct for the data obtained from the simulation (see section \ref{ssec:validation}). A likelihood-ratio test is then performed to see if the resulting observations could have been an extreme sample from the true process (see section \ref{ssec:hyptest} and table \ref{tab:LRtest}). 

The results indicate that market mechanisms and practical considerations have significantly altered the process and can indeed affect our ability to infer the underlying process, as evidenced in figures \ref{fig:alpha distortion} and \ref{fig:beta distortion}. Our findings suggest that when considering market models, such as agent-based models at least in the setting of continuous double auctions, it is prudent to account for practical considerations imposed by market rules, mechanism and asynchronicity. It may be inadequate to argue their impacts away using additional sources of noise or systematic model parameter uncertainties.

From a pragmatic perspective this suggests that focusing the modelling on event times separate from the event characteristics can lead to ambiguities in model interpretation. This suggests that more realistic market models need to combine event time models directly with processes that generate their sizes, order types and interactions between orders and their environment.

\section*{Reproducibility}

CoinTossX \href{https://github.com/dharmeshsing/CoinTossX/tree/v1.1.0}{v1.1.0} was used in this work. All code and instructions for replication can be found at: \newline \href{https://github.com/IvanJericevich/IJPCTG-HawkesCoinTossX}{https://github.com/IvanJericevich/IJPCTG-HawkesCoinTossX}. 

\section*{Acknowledgements \label{sec:Acknowledgements}}
We would like to thank Etienne Pienaar, Birgit Erni, Dharmesh Sing and Dieter Hendricks for various fruitful discussions and constructive feedback.

\section*{Funding \label{sec:Funding}}
This research did not receive any specific grant from funding agencies in the public, commercial, or not-for-profit sectors.

\balance
\bibliographystyle{elsarticle-harv}
\bibliography{HMS-CoinTossX}

%-----------------------------------------------------
%	APPENDIX
%-----------------------------------------------------
\onecolumn
\appendix
\section{Pseudocode}

\begin{algorithm} % Made this pseudocode 1
    \small
    \vspace{0pt}
    \SetAlgoLined
    \DontPrintSemicolon
    \KwIn{Event relative times $t_i$, volume $v_{t_i}$, type, classification, and side generated from a 10-variate Hawkes process}
    Login a single client to CoinTossX to trade in a single security \;
    Initialize the previous best bid $b_{t_0}$ and best ask $a_{t_0}$ at a price level of 1000 where $t_0 = 0$ \;
    Convert all order submission relative times to clock time ($t$): $t_i \ \forall i \in \{1, \hdots, N\}$ \;
    Submit the first order at a starting price of 1000 \;
    \ForEach{$order_{t_i}$ with $i \in 2, \hdots, N$}{
        Retrieve the best bid $b_{t_i}$ and best ask $a_{t_i}$ from the market data feed \;
        \uIf{$order_{t_i}$ is a limit order}{
            \eIf{The LOB is empty: $b_{t_i} == 0$ and $a_{t_i} == 0$}{
                \lIf{$order_{t_i}$ is a bid}{Place the limit price $p_{t_i}$ according to the rules in algorithm \ref{algo:model1} or \ref{algo:model2} with input parameters $0$ for the best bid and $a_{t_{i - 1}}$ for the previous best ask}\lElse{Place the limit price $p_{t_i}$ according to the rules in algorithms \ref{algo:model1} or \ref{algo:model2} with input parameters $b_{t_{i - 1}}$ for the previous best bid and $0$ for the previous best ask}
            }{
                Place the limit price $p_{t_i}$ according to the rules in algorithm \ref{algo:model1} or \ref{algo:model2} with input parameters $b_{t_i}$ for the best bid and $a_{t_i}$ for the best ask
            }
            \lIf{If the current time is less than the arrival time of $order_{t_i}$: $t < t_i$}{Wait for the remaining duration until the arrival time to submit the limit order at price $p_{t_i}$ with volume $v_{t_i}$}\lElse{The algorithm is lagging behind the times specified by the Hawkes process, so submit the limit order at price $p_{t_i}$ with volume $v_{t_i}$ immediately}
        }\uElseIf{event $i$ is a market order and the contra side is non-empty}{
            \lIf{If the current time is less than the arrival time of $order_{t_i}$: $t < t_i$}{Wait for the remaining duration until the arrival time to submit the market order with volume $v_{t_i}$}\lElse{The algorithm is lagging behind the times specified by the Hawkes process, so submit the market order with volume $v_{t_i}$ immediately}
        }\uElseIf{event $i$ is a cancel order and the corresponding side of the LOB is non-empty}{
            Retrieve a LOB snapshot from the market data feed \;
            \eIf{$event_i$ is an aggressive order}{
                Extract the order IDs from the corresponding LOB side that are active at level 1: $\mathcal{L}^1_{t_i}$ \;
                Select an order $o$ by randomly sampling from $\mathcal{L}^1_{t_i}$ \;
            }{
                Extract the order IDs from the corresponding LOB that are active at levels $> 2$: $\mathcal{L}^{2+}_{t_i}$ \;
                Select an order $o$ by randomly sampling from $\mathcal{L}^{2+}_{t_i}$ \;
            }
            \lIf{If the current time is less than the arrival time of $order_{t_i}$: $t < t_i$}{Wait for the remaining duration until the arrival time to cancel order $o$}\lElse{The algorithm is lagging behind the times specified by the Hawkes process, so cancel order $o$ immediately}
        }
    }
    Logout the client from CoinTossX \;
    \caption{Basic logic for the submission of Hawkes-generated orders to CoinTossX \label{algo:injectsimulation}}
\end{algorithm}

\begin{minipage}[t]{.5\linewidth}
\begin{algorithm}[H]
    \small
    \vspace{0pt}
    \SetAlgoLined
    \DontPrintSemicolon
    \KwIn{The best bid $b_i$ prior to the submission of the $i$th event, the best ask $a_i$ prior to the submission of the $i$th event, and the $i$th order $\text{order}_i$}
    \KwResult{Limit price for the $i$th order}
    Randomly sample $u \in \{1, \hdots, 10\}$ with equal probability\;
    \eIf{$\text{order}_i$ is a bid}{
        \eIf{$\text{order}_i$ is passive}{
            \eIf{Contra side is empty $a_i == 0$}{
                $p_i = b_i - 1$ \;
            }{
                \lIf{The bid side of the LOB is empty $b_i == 0$}{Place the price a random number of ticks opposite the contra-best $p_i = a_i - u$}\lElse{$p_i = b_t - 1$}
            }
        }{
            \lIf{The bid side of the LOB is empty $b_i == 0$}{Place the price a random number of ticks opposite the contra-best $p_i = a_t - u$}\lElse{Improve the best by 1 tick $p_i = b_t + 1$}
        }
    }{
        \eIf{$\text{event}_i$ is passive}{
            \eIf{Contra side is empty $b_i == 0$}{
                $p_i = a_t + 1$ \;
            }{
                \lIf{The ask side of the LOB is empty $a_i == 0$}{Place the price a random number of ticks opposite the contra-best $p_i = b_t + u$}\lElse{$p_i = a_t + 1$ }
            }
        }{
            \lIf{The ask side of the LOB is empty $a_i == 0$}{Place the price a random number of ticks opposite the contra-best $p_i = b_t + u$}\lElse{Improve the best by 1 tick $p_i = a_t - 1$}
        }
    }
    \caption{Limit price placement in model 1 \label{algo:model1}}
\end{algorithm}
\end{minipage}
\begin{minipage}[t]{.5\linewidth}
\begin{algorithm}[H]
    \small
    \vspace{0pt}
    \SetAlgoLined
    \DontPrintSemicolon
    \KwIn{The best bid $b_i$ prior to the submission of the $i$th event, the best ask $a_i$ prior to the submission of the $i$th event, and the $i$th order $\text{order}_i$}
    \KwResult{Limit price for the $i$th order}
    Randomly sample $u \in \{1, \hdots, 10\}$ with equal probability \;
    \eIf{$\text{order}_i$ is a bid}{
        \eIf{$\text{order}_i$ is passive}{
            \eIf{Contra side is empty $a_i == 0$}{
                $p_i = b_i - 1$ \;
            }{
                \lIf{The bid side of the LOB is empty $b_i == 0$}{Place the price a random number of ticks opposite the contra-best $p_i = a_i - u$}\lElse{$p_i = b_t - 1$}
            }
        }{
            \eIf{The bid side of the LOB is empty $b_i == 0$}{
                Place the price a random number of ticks opposite the contra-best $p_i = a_t - u$ \;
            }{
                Calculate the spread as $s_i = |a_i - b_i|$ \;
                \lIf{The spread is greater than 1 tick $s_i > 1$}{Improve the best by 1 tick $p_i = b_t + 1$}\lElse{Place the price at the best $p_i = b_t$}
            }
        }
    }{
        \eIf{$\text{event}_i$ is passive}{
            \eIf{Contra side is empty $b_i == 0$}{
                $p_i = a_t + 1$ \;
            }{
                \lIf{The ask side of the LOB is empty $a_i == 0$}{Place the price a random number of ticks opposite the contra-best $p_i = b_t + u$}\lElse{$p_i = a_t + 1$ }
            }
        }{
            \eIf{The ask side of the LOB is empty $a_i == 0$}{
                Place the price a random number of ticks opposite the contra-best $p_i = b_t + u$ \;
            }{
                Calculate the spread as $s_i = |a_i - b_i|$ \;
                \lIf{The spread is greater than 1 tick $s_i > 1$}{Improve the best by 1 tick $p_i = a_t - 1$}\lElse{Place the price at the best $p_i = a_t$}
            }
        }
    }
    \caption{Limit price placement in model 2 \label{algo:model2}}
\end{algorithm}
\end{minipage}

\newpage

\section{Calibration Likelihood} \label{app:likelihood}
Following \citet{TP2012} the log-likelihood for the Hawkes process is defined as:
\begin{equation}\label{hawkes:eq:5}
    \ln \mathcal{L}\left( \left\{ N\left(t\right) \right\}_{t\leq T}  \right) = \sum_{m=1}^M \ln \mathcal{L}^m\left( \left\{ N_m\left(t\right) \right\}_{t\leq T} \right),
\end{equation}
with each term defined as:
\begin{equation}\label{hawkes:eq:6}
    \begin{aligned}
        \ln \mathcal{L}^m\left( \left\{ N_m\left(t\right) \right\}_{t\leq T} \right) 
        &= \int_0^T (1 - \lambda^m(s)) ds + \int_0^T \ln \lambda^m(s) dN_m(s).
    \end{aligned}
\end{equation}
Using the exponential kernel, \cref{hawkes:eq:6} can be computed as:
\begin{equation}\label{hawkes:eq:7}
\begin{aligned}
    \ln \mathcal{L}^m\left( \left\{ N_m\left(t\right) \right\}_{t\leq T} \right) 
    &= T - \int_0^T \lambda^m(s) ds + \sum_{i:t_i \leq T} \mathbbm{1}_{\{Z_i = m\}} \ln \bigg[ \mu^m + \sum_{n=1}^M \sum_{t_k^n < t_i} \alpha^{mn} e^{-\beta^{mn}(t_i - t_k^n)} \bigg].
\end{aligned}
\end{equation}
We streamline \cref{hawkes:eq:7} using the Markov property
\begin{equation}\label{hawkes:eq:8}
    R^{mn}(l) 
    = \sum_{t_k^n < t_l^m} e^{-\beta^{mn}(t_l^m - t_k^n)}   %\\
    =\begin{cases} 
      e^{-\beta^{mn}(t_l^m - t_{l-1}^m)} R^{mn}(l-1) +\sum_{t_{l-1}^m \leq t_k^n < t_{l}^m} e^{-\beta^{mn}(t_l^m - t_k^n)} & \text{if } m\neq n, \\
      e^{-\beta^{mn}(t_l^m - t_{l-1}^m)} \left( 1 + R^{mn}(l-1) \right) & \text{if } m = n,
  \end{cases}
\end{equation}
subject to $R^{mn}(0) = 0$. Here the exponential kernel can be used to obtain the well known recursion relation \cref{hawkes:eq:8}. This is because with the exponential kernel $\lambda(t)$ can be recast into Markovian form so that we do not need to keep track of the entire history. This reduces the computation of the likelihood from $\mathcal{O}(n^2M)$ to $\mathcal{O}(nM)$, where $n$ is the total number of arrival times.

Including \cref{hawkes:eq:8} in \cref{hawkes:eq:7} gives us:
\begin{equation}\label{hawkes:eq:9}
\begin{aligned}
    \ln \mathcal{L}^m\left( \left\{ N_m\left(t\right) \right\}_{t\leq T} \right) 
    &= T - \int_0^T \lambda^m(s) ds - \sum_{i:t_i \leq T} \sum_{m=1}^M \frac{\alpha^{mn}}{\beta^{mn}} \left( 1 - e^{-\beta^{mn} (T - t_i)} \right)   \\
    &+ \sum_{l: t_l^m \leq T} \ln \left[ \lambda^m(t_l^m) + \sum_{n=1}^M \alpha^{mn} R^{mn}(l) \right].
\end{aligned}
\end{equation}
The log-likelihood can then be maximised
\begin{equation}\label{hawkes:eq:10}
    \hat{\theta} = \underset{\theta}{\text{argmax }} \ln \mathcal{L}\left( \theta \right), 
\end{equation}
where $\theta$ refers to the parameters of the Hawkes process which include $(\mu^m, \alpha^{mn}, \beta^{mn})$, depending on the specification of the model. Here we do not enforce a specific design on the matrices $\boldsymbol{\alpha}$ or $\boldsymbol{\beta}$ and allow the calibration to determine the structure of the matrices. Therefore, we have a total of 210 parameters.

Our optimisation of \cref{hawkes:eq:10} employs the gradient descent method L-BFGS from the optimisation package ``Optim'' \cite{MRA2018}. The method takes steps according to
\begin{equation}\label{eq:10}
    \theta_{h+1} = \theta_{h} - \kappa P^{-1} \nabla \ln \mathcal{L} \left( \theta_{h} \right),
\end{equation}
where $\kappa$ is a scalar chosen by a linesearch algorithm to ensure that each step gives sufficient descent, $P$ is an approximation to the Hessian using the latest $m$ steps and $\nabla \ln \mathcal{L}$ is the first derivative of our log-likelihood found through numerical approximations using the package ``ForwardDiff'' \cite{RLP2016}. The search begins at the true parameter values $\theta_{\text{true}}$ defined in \cref{subsec:setup}.

To the best of our knowledge, closed form solutions for the gradients and Hessian have not been found for a multivariate Hawkes process. Closed form solutions for the univariate case can be found in \cite{Ozaki1979}.

\section{Calibrated parameters}\label{app:param}
\subsection{Visualisation}

\begin{figure*}[!h]
    \centering
    \subfloat[$\bm{\mu}$ distortions: Reference model]{\includegraphics[width=.33\textwidth]{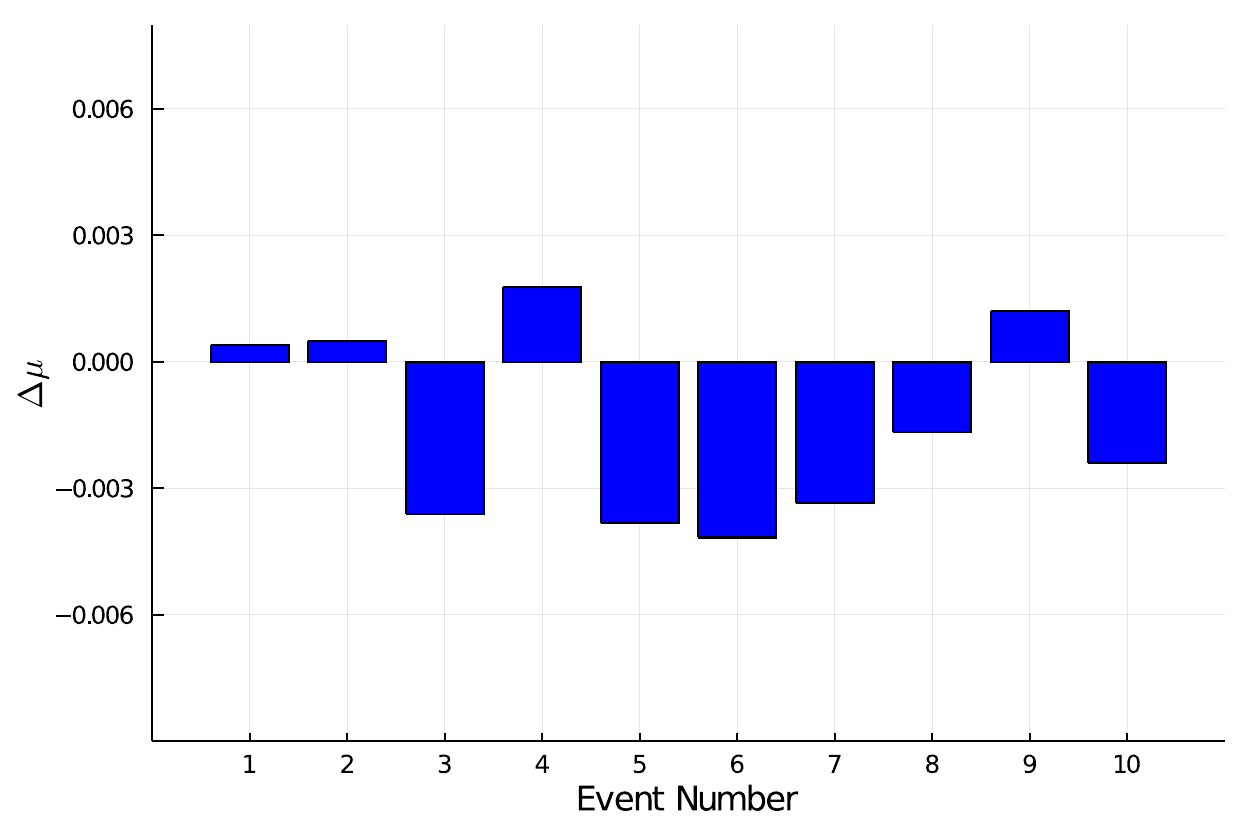}}
    \subfloat[$\bm{\mu}$ distortions: Model 1]{\includegraphics[width=.33\textwidth]{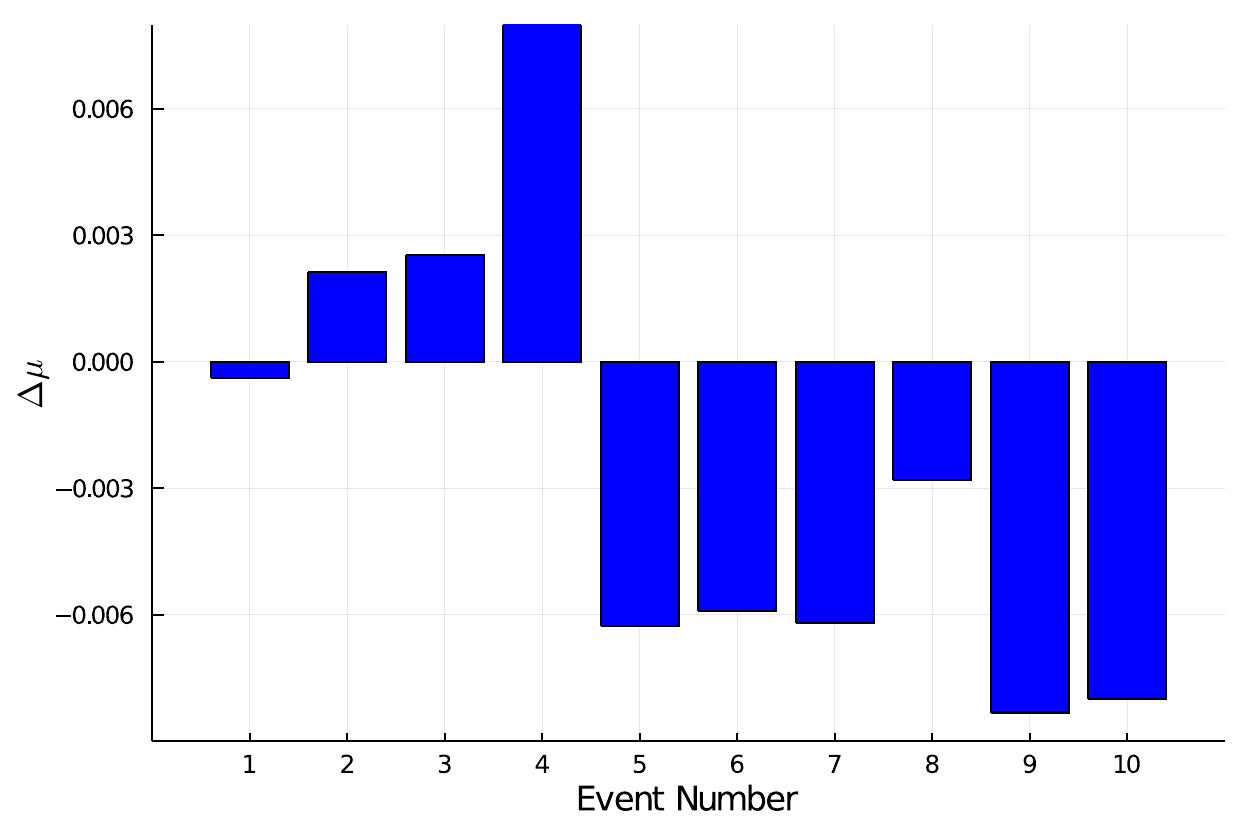}}
    \subfloat[$\bm{\mu}$ distortions: Model 2]{\includegraphics[width=.33\textwidth]{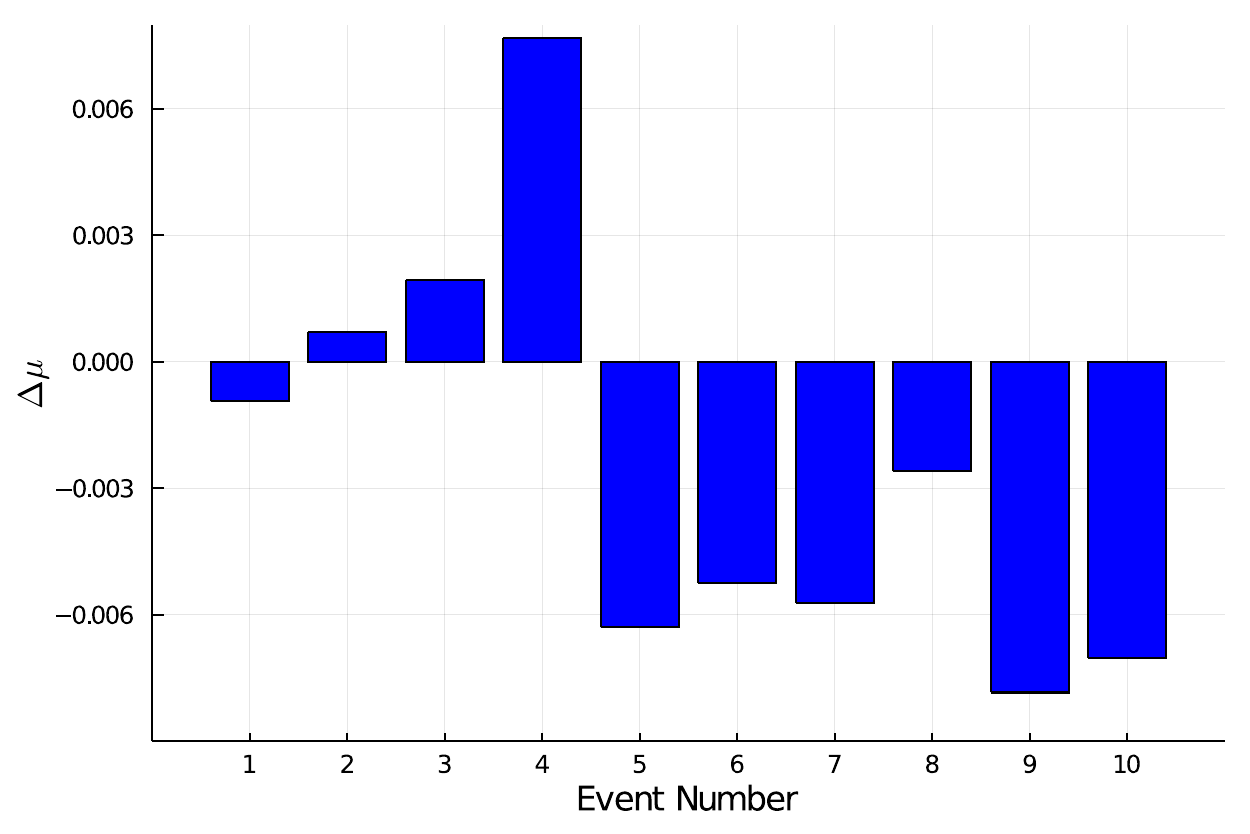}}
    \caption{Parameter distortions in the baseline intensity $\Delta {\bm {\mu}}$ between the calibrated parameters and the true parameters from the underlying process. The baseline intensities in model 1 and model 2 are found to have similar distortion magnitudes. The largest of these came from aggressive asks (event 4), passive LOs (events 5 and 6) and passive cancels (events 9 and 10).}
    \label{fig:mu distortion}
\end{figure*}

We visualise the deviation between parameters from the calibration against the true parameters from the underlying process. First, the deviation between the baseline intensities $\Delta \bm{\mu} = \hat{\bm{\mu}} - \bm{\mu}_{\text{true}}$ are visualised using bar plots in figure~\ref{fig:mu distortion}. We see that the deviations are not particularly large.

\begin{figure*}[!h]
    \centering
    \subfloat[$\bm{\alpha}$ distortions: Reference model]{\includegraphics[width=.33\textwidth]{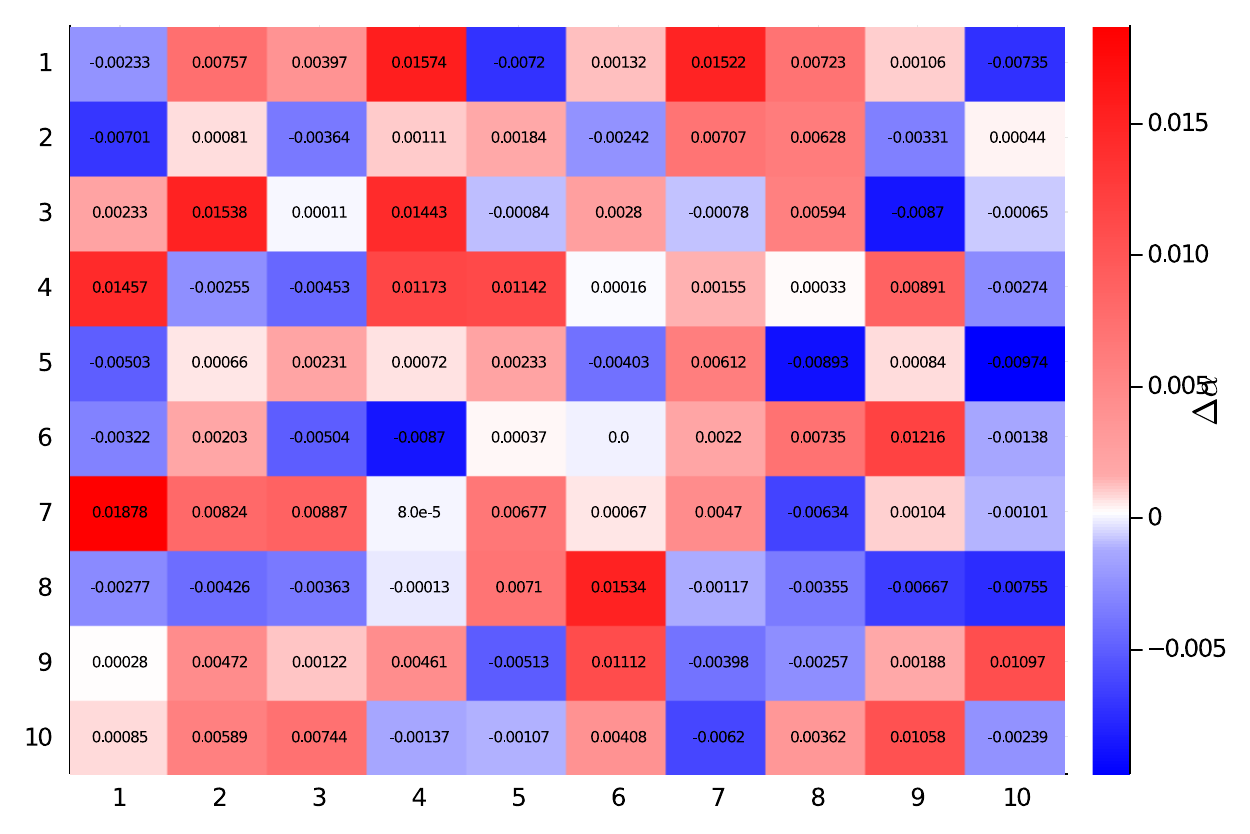}}
    \subfloat[${\bm{\alpha}}$ distortions: Model 1]{\includegraphics[width=.33\textwidth]{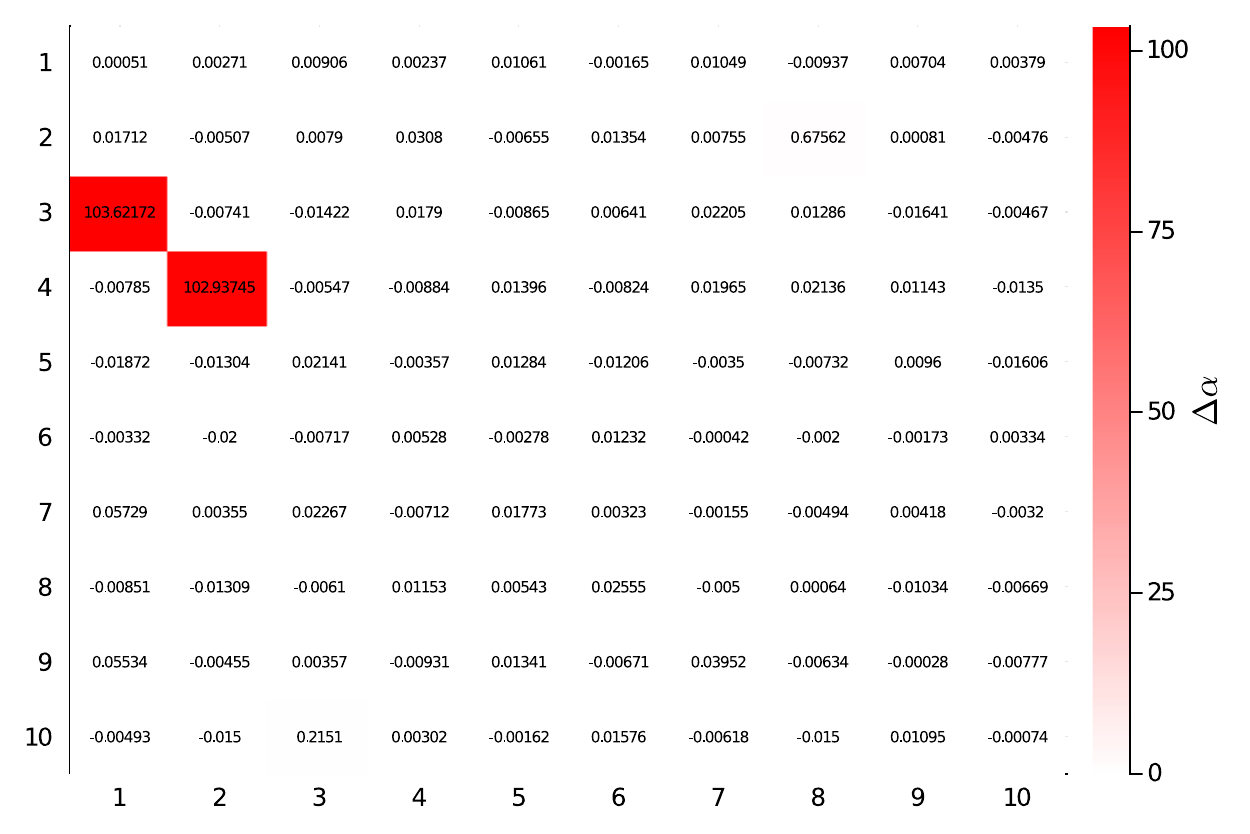}}
    \subfloat[${\bm{\alpha}}$ distortions: Model 2]{\includegraphics[width=.33\textwidth]{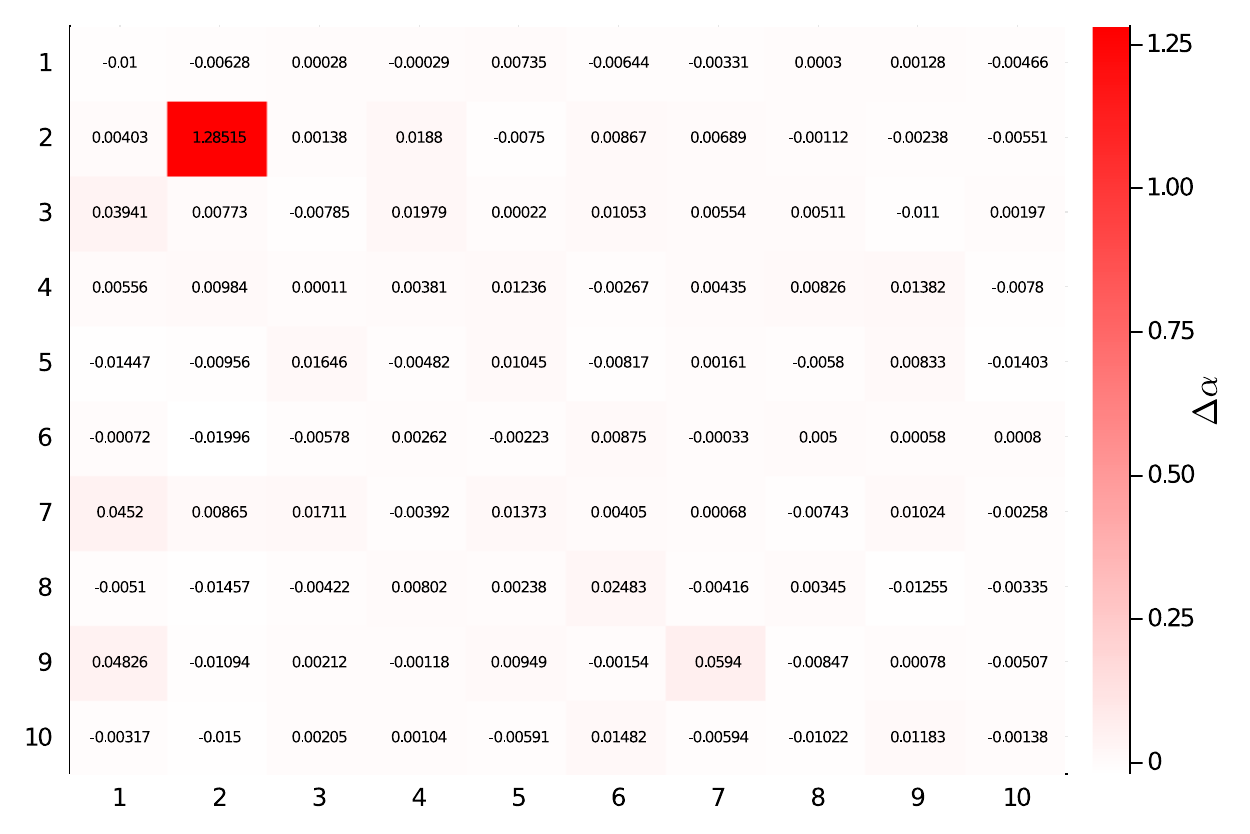}}
    \caption{Parameter distortions in the excitation matrix $\Delta {\bm{\alpha}}$ between the calibrated parameters and the true parameters from the underlying process.}
    \label{fig:alpha distortion}
\end{figure*}

\begin{figure*}[!h]
    \centering
    \subfloat[$\bm{\beta}$ distortions: Reference model]{\includegraphics[width=.33\textwidth]{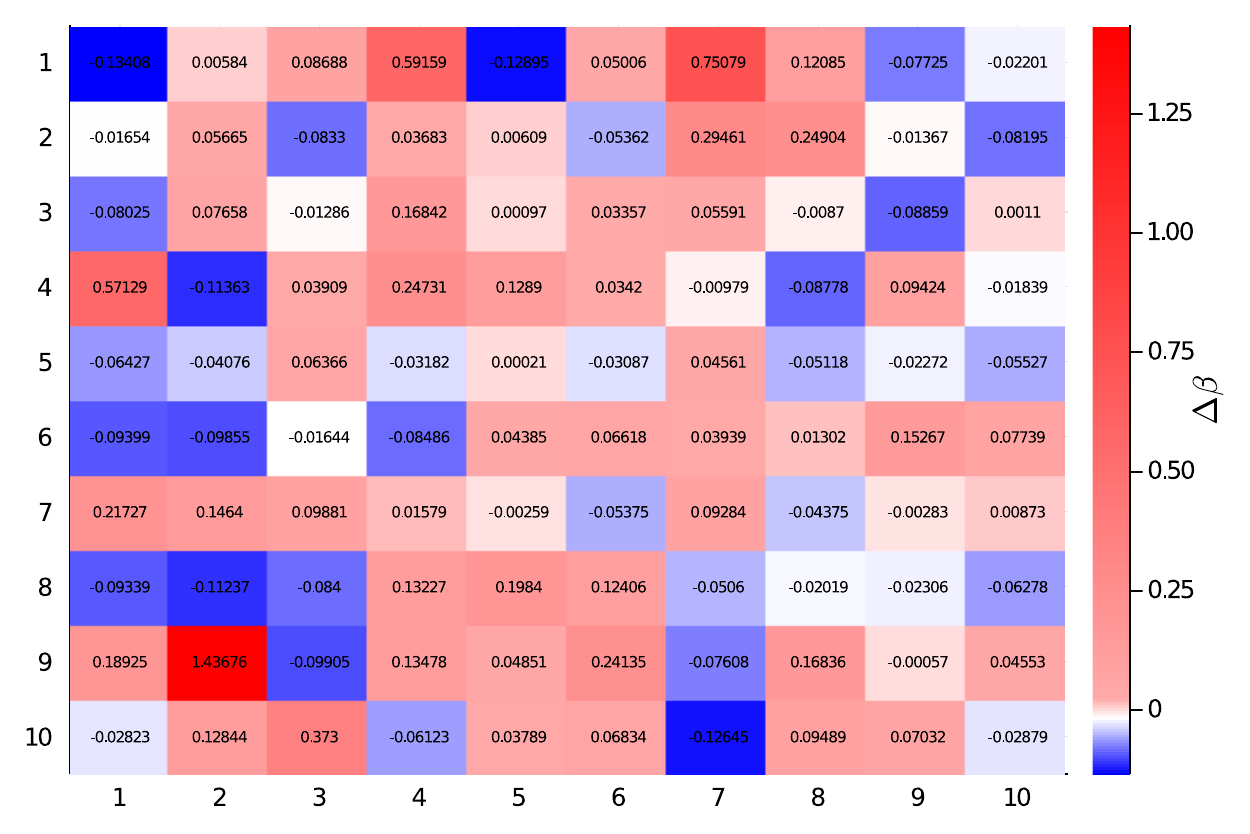}}
    \subfloat[$\bm{\beta}$ distortions: Model 1]{\includegraphics[width=.33\textwidth]{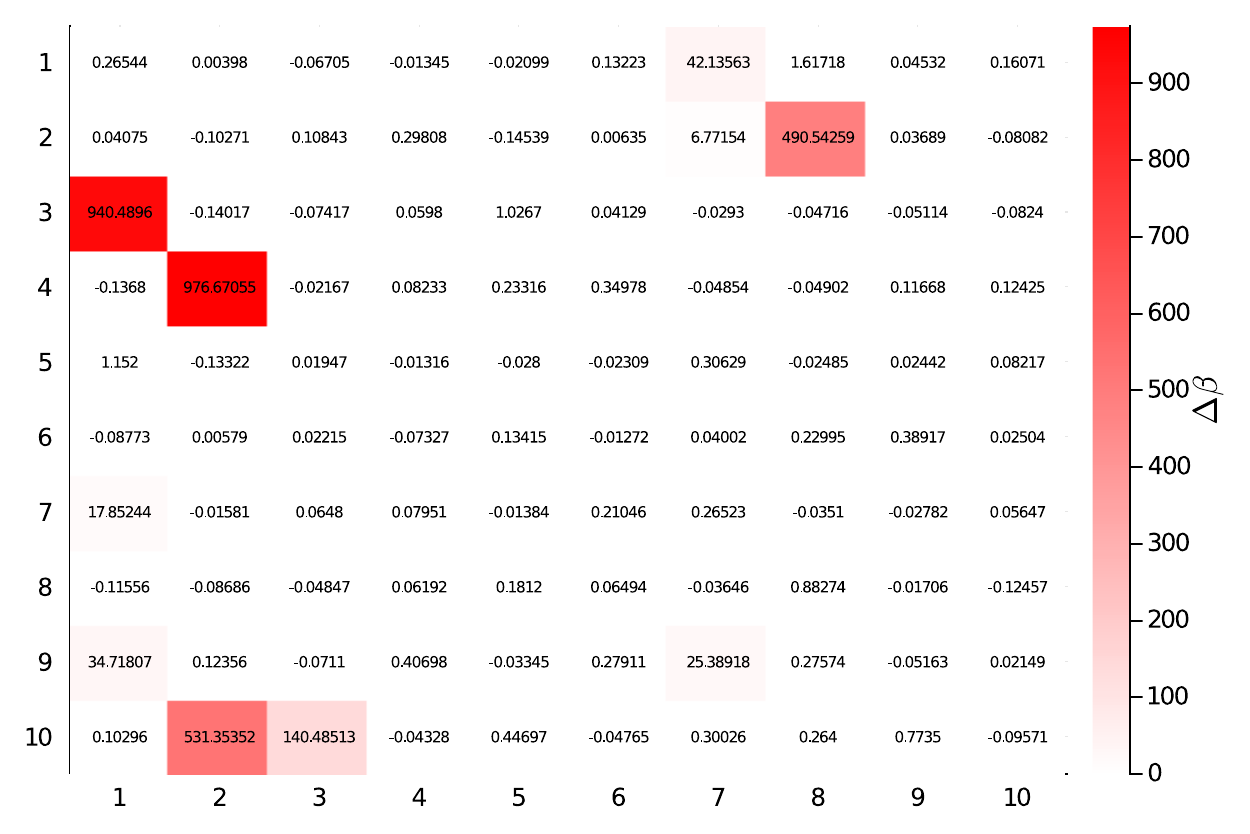}}
    \subfloat[$\bm{\beta}$ distortions: Model 2 ]{\includegraphics[width=.33\textwidth]{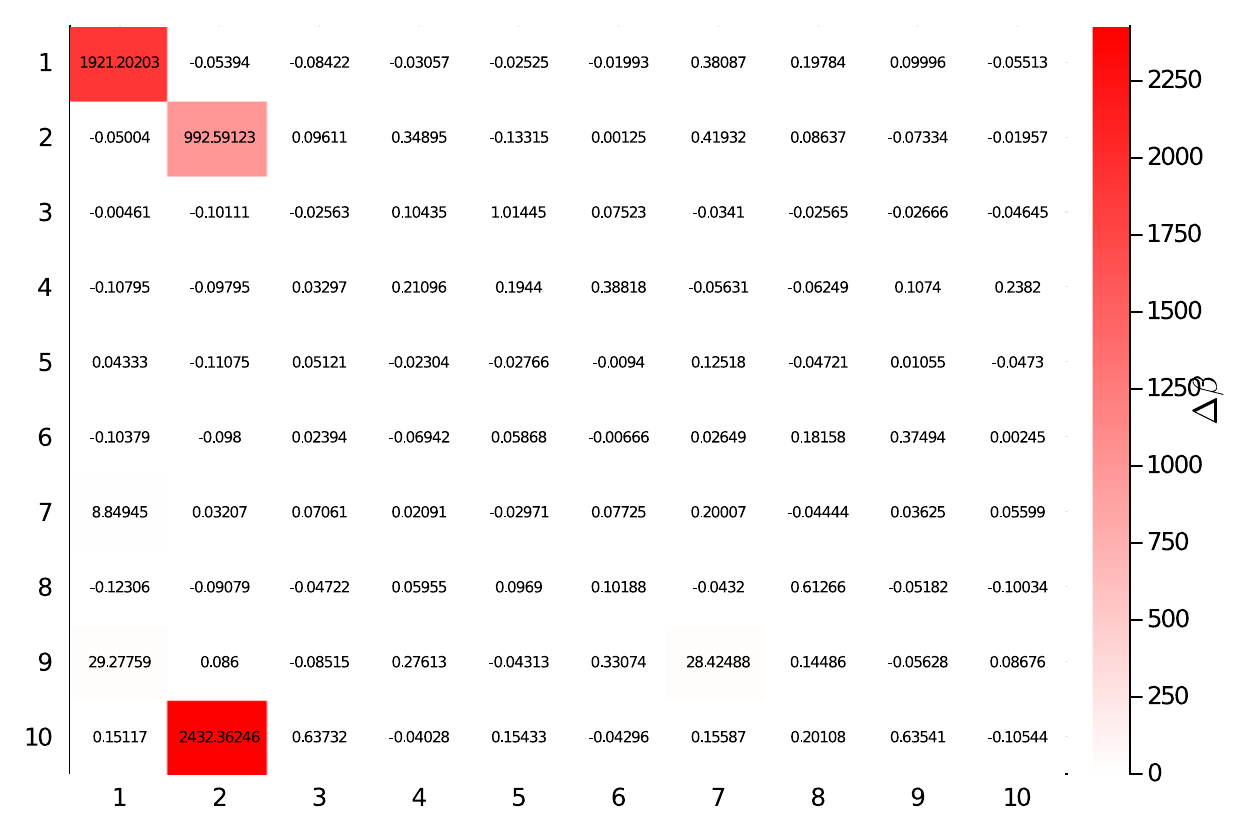}}
    \caption{Parameter distortions in the decay matrix $\Delta {\bm{\beta}}$ between the calibrated parameters and the true parameters from the underlying process.}
    \label{fig:beta distortion}
\end{figure*}

Second, the distortion between $\Delta \bm{\alpha} = \hat{\bm{\alpha}} - \bm{\alpha}_{\text{true}}$ and $\Delta \bm{\beta} = \hat{\bm{\beta}} - \bm{\beta}_{\text{true}}$ are visualised using heatmaps in figures~\ref{fig:alpha distortion} and \ref{fig:beta distortion} respectively. The deviations in the reference model are minimal whereas the deviations in model 1 and 2 are relatively large. We will highlight and discuss a few notable changes that can be explained through the various market mechanics. 

Looking at $\alpha^{3,1}, \alpha^{4,2}, \beta^{3,1}$ and $\beta^{4,2}$ for model 1, these deviations can be attributed towards the additional spawned event caused by aggressive limit orders crossing the book. This is because when an aggressive limit order crosses the book, it first becomes a market order and if it has a larger volume than the contra side then an additional limit order will be spawned as discussed in section \ref{ssec:mech}. Therefore, to capture an aggressive limit order that spawns immediately after a market order (which was converted), $\alpha^{3,1}$ and $\alpha^{4,2}$ has to significantly increase to ensure that an event is spawned immediately and the event will become an aggressive limit order (from the same side as the market order). However, this increased intensity should only last for a very short while since only one event is spawned. Thus to accommodate this, the rate of decay in $\beta^{3,1}$ and $\beta^{4,2}$ must also significantly increase to avoid spawning too many additional events.

Looking at $\beta^{1,1}$ and $\beta^{2,2}$ for model 2, these deviations can be attributed towards the dropping of events. This is because too many consecutive market orders coming through from the same side will deplete liquidity on the contra side and thus emptying the LOB. Once the contra side of the LOB is empty, any additional market orders will not go through. Therefore to capture this, the rate of decay must increase to capture the dropped orders.

\begin{figure*}[!h]
    \centering
    \subfloat[$\bm{\Gamma}$ distortions: Reference model]{\includegraphics[width=.33\textwidth]{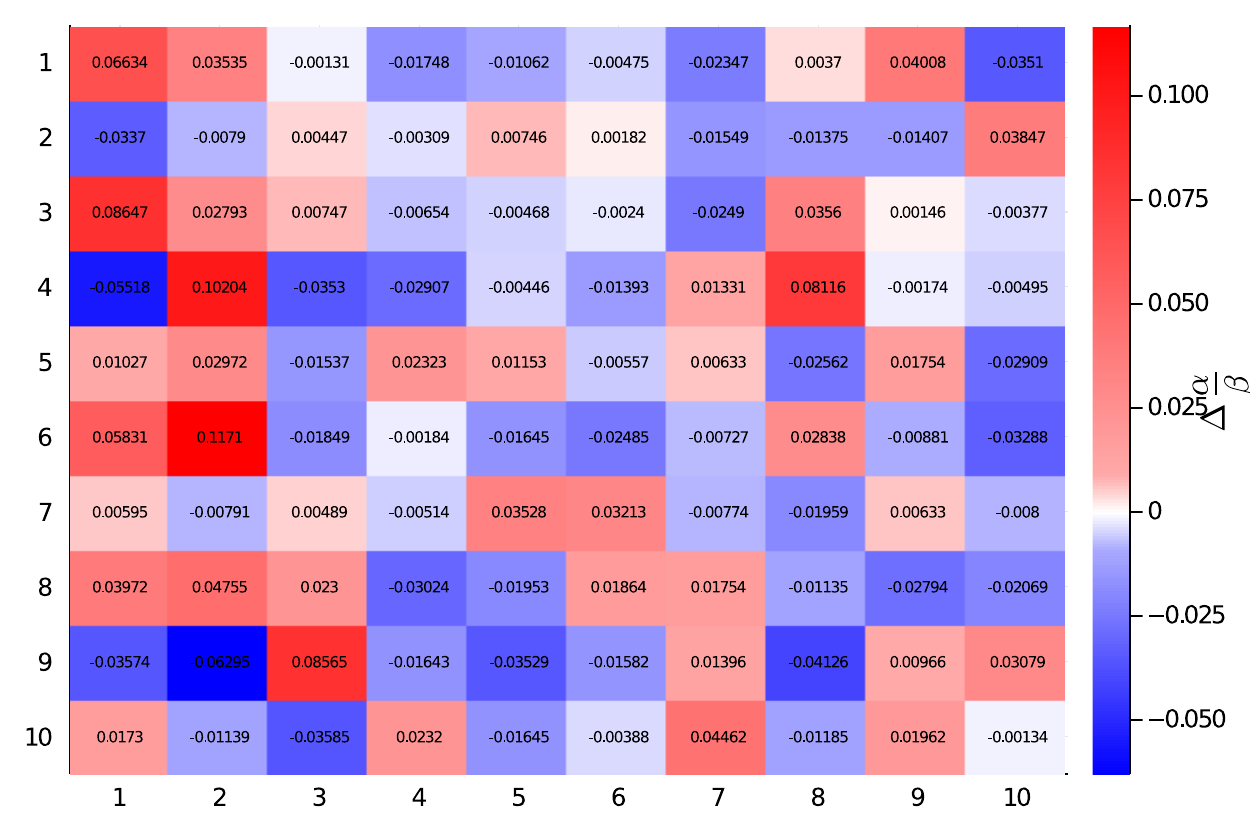}}
    \subfloat[$\bm{\Gamma}$ distortions: Model 1]{\includegraphics[width=.33\textwidth]{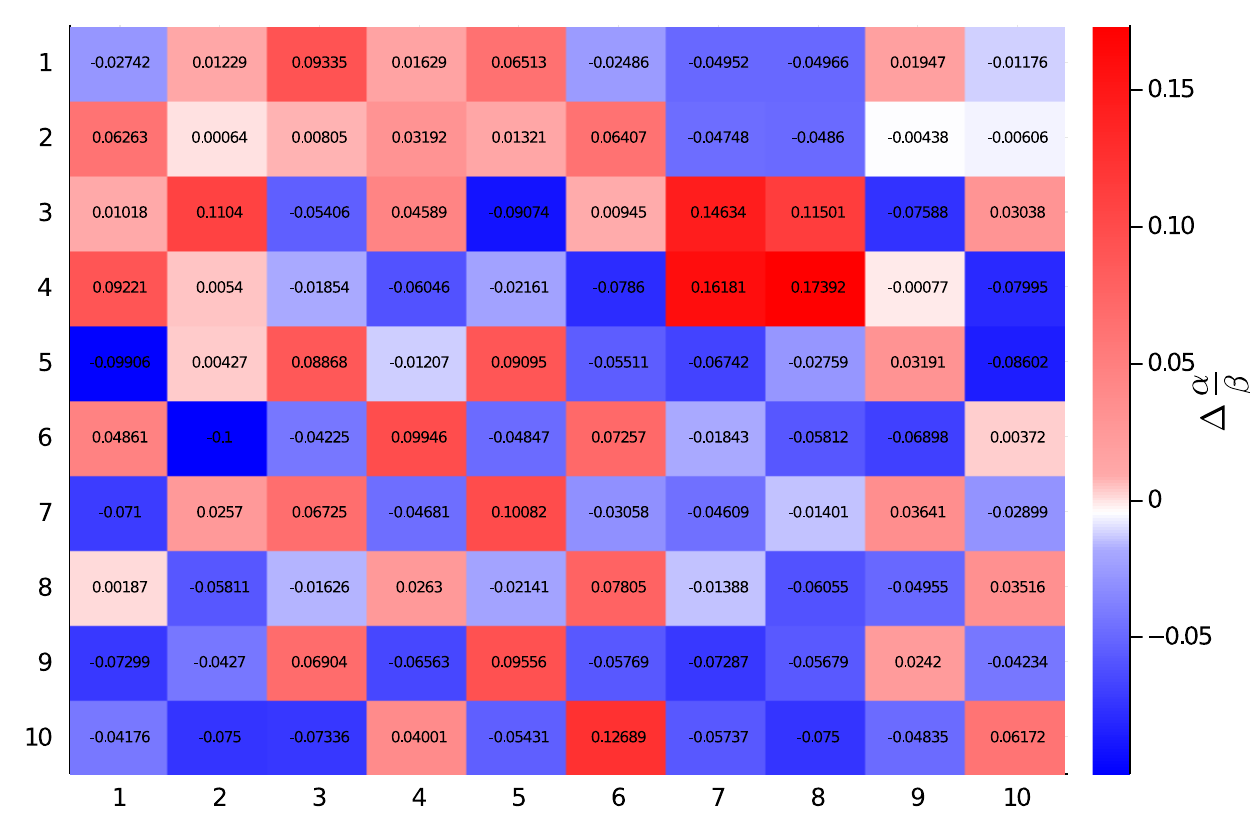}}
    \subfloat[$\bm{\Gamma}$ distortions: Model 2 ]{\includegraphics[width=.33\textwidth]{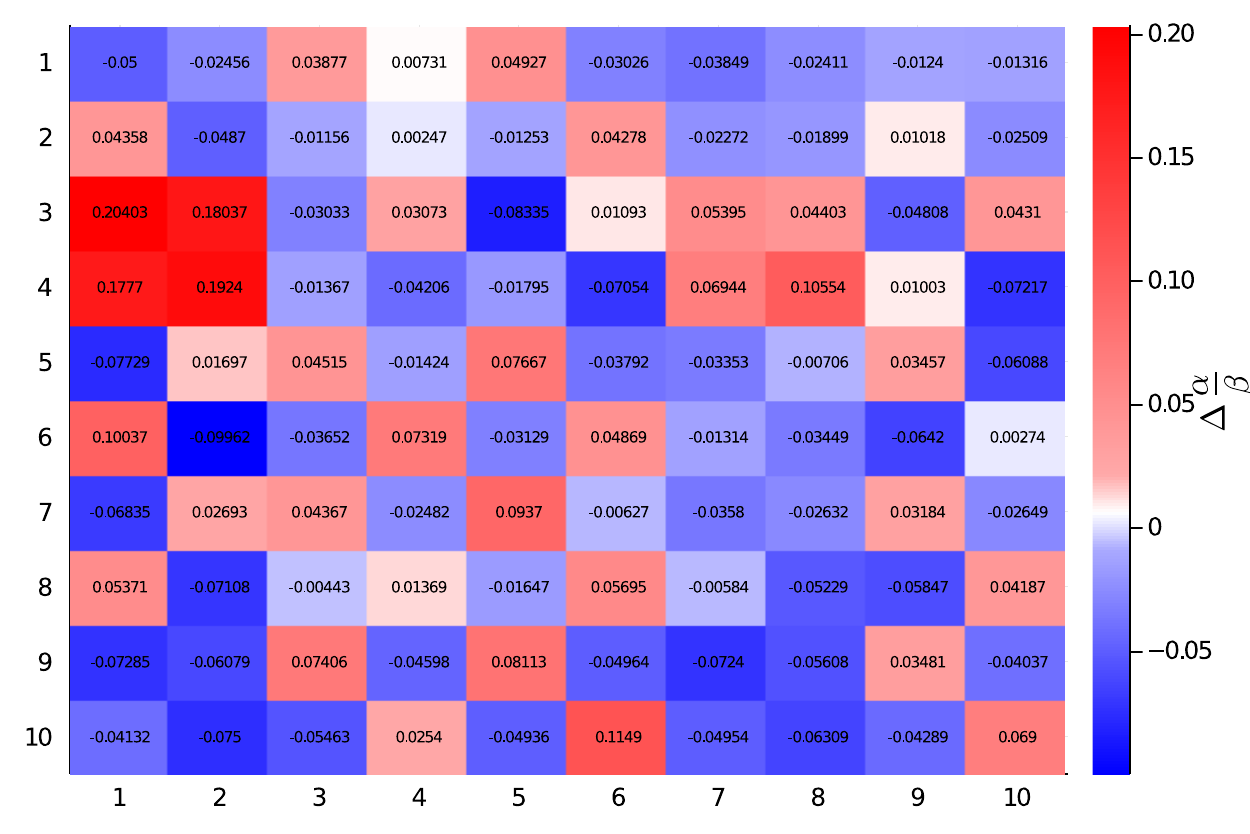}}
    \caption{Branching ratio distortions $\Delta {\bm {\Gamma}}$ between the calibrated parameters and the true parameters from the underlying process. Despite the many distortions, the expected average number of events remain fairly well balanced.}
    \label{fig:branching ratio distortion}
\end{figure*}

Next, the distortion of the branching ratios $\Delta \bm{\Gamma} = \hat{\bm{\Gamma}} - \bm{\Gamma}_{\text{true}}$ are visualised in figure \ref{fig:branching ratio distortion}. In model 1 the events whose branching ratios are most significantly increased are that of aggressive cancels that spawn aggressive LOs on both sides of the LOB (the effect of events 7 and 8 on events 3 and 4). The reason for the increase in expected number of events in this case is that cancel orders often empty the LOB, resulting in any LO occuring thereafter being classified as aggressive irrespective of whether the original order was passive or not. % Related to this is the decrease in the xpected number of events that result from buy trades spawning passive buy LOS and sell trades spawning passive sell LOs

Similarly, for model 2 the most significant increase in branching ratios occur for market orders that spawn aggressive limit orders. This can be explained by the fact that trades emptying the LOB will cause limit orders occurring thereafter to always become aggressive --- thus resulting in greater numbers of offspring from those events. %Related to this, the most significant decrease occurs for sell market orders that spawn passive sell limit orders. This is also because passive limit orders occuring after a large  in an empty LOB become aggressive LOs

\begin{figure*}[htb]
    \centering
    \subfloat[Excitations on event type 2 \label{subfig:if1-2}
    ]{\includegraphics[width=.5\textwidth]{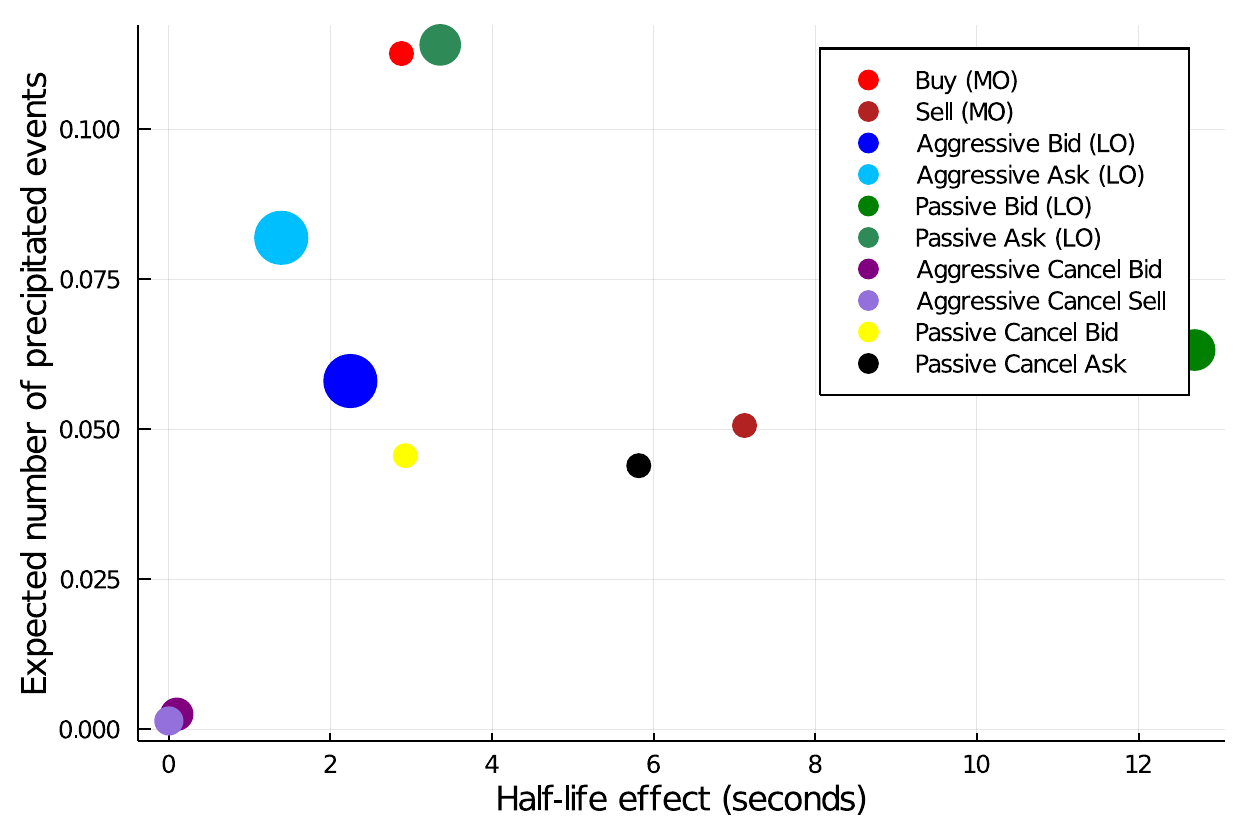}}
    \subfloat[Excitations on event type 3 \label{subfig:if1-3}
    ]{\includegraphics[width=.5\textwidth]{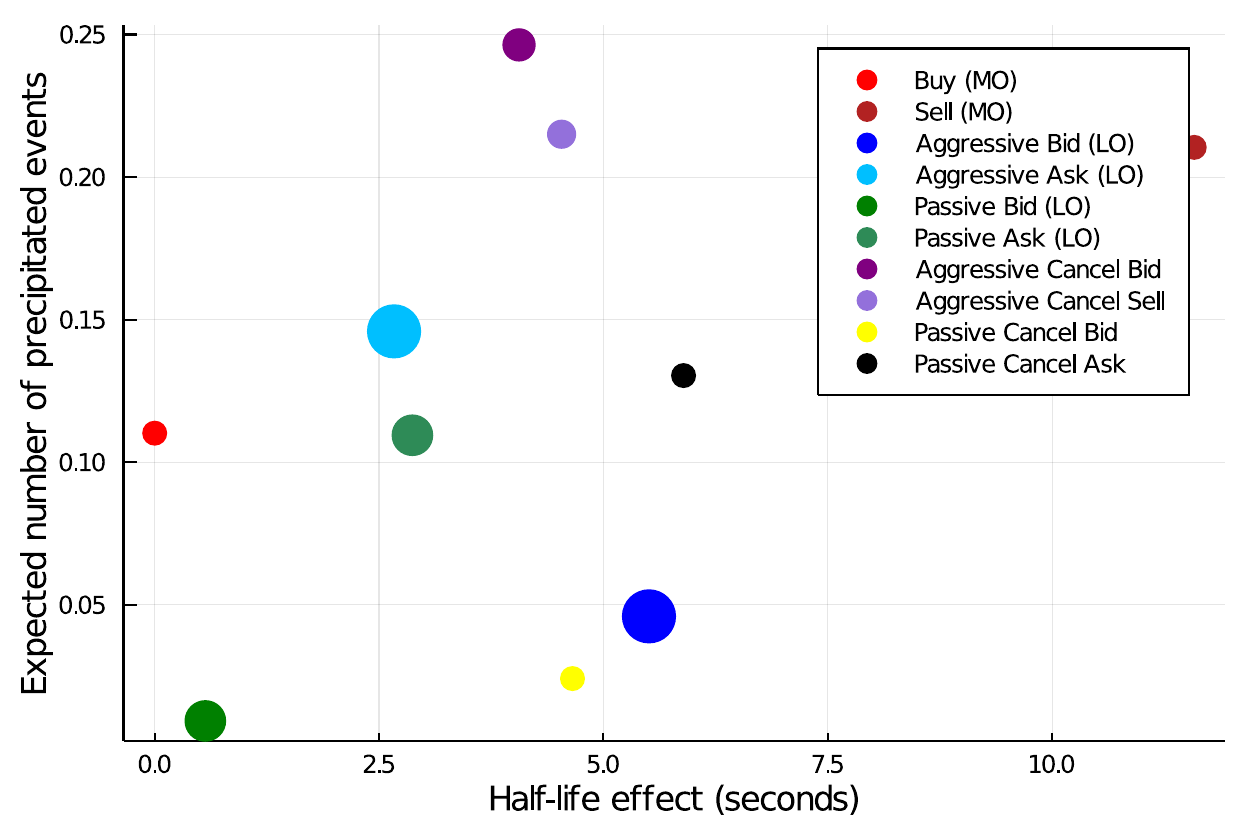}} \\
    \subfloat[Excitations on event type 4 \label{subfig:if1-4}
    ]{\includegraphics[width=.5\textwidth]{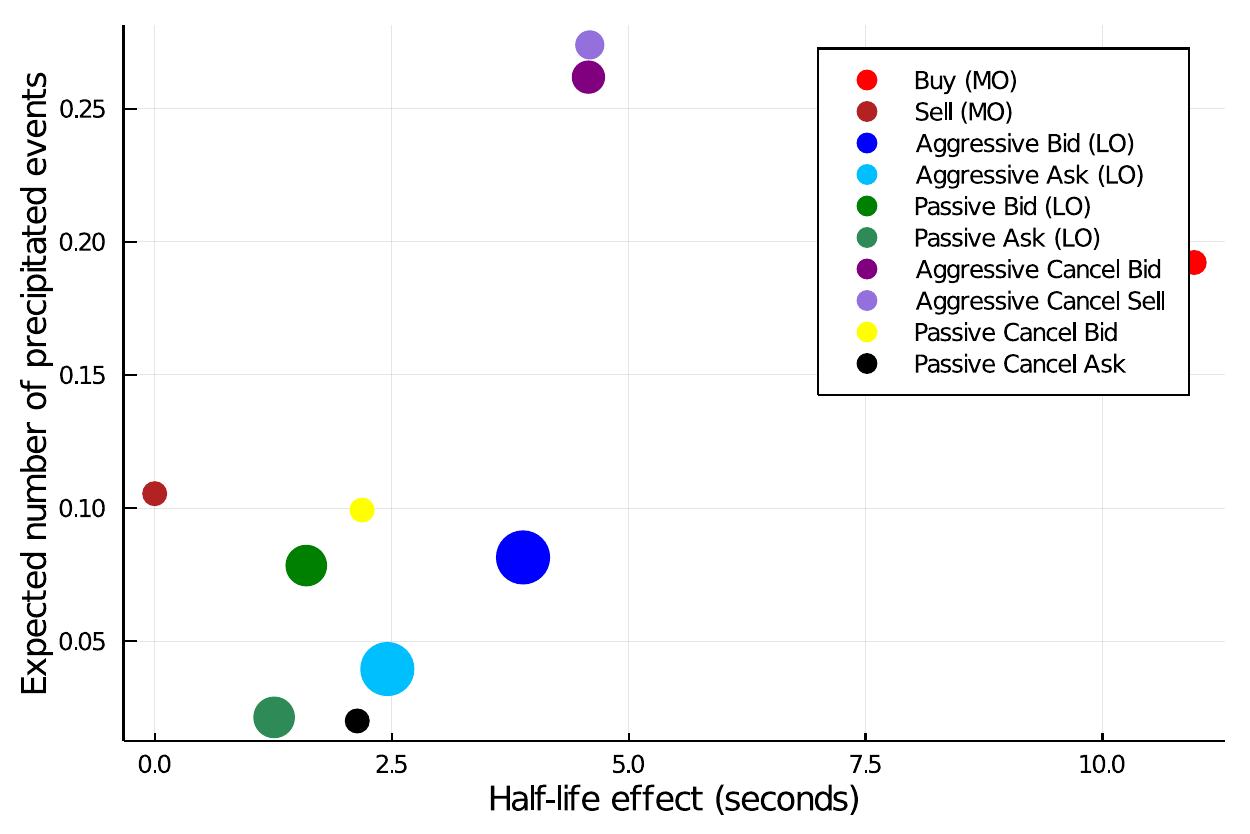}}
    \subfloat[Excitations on event type 10 \label{subfig:if1-10}
    ]{\includegraphics[width=.5\textwidth]{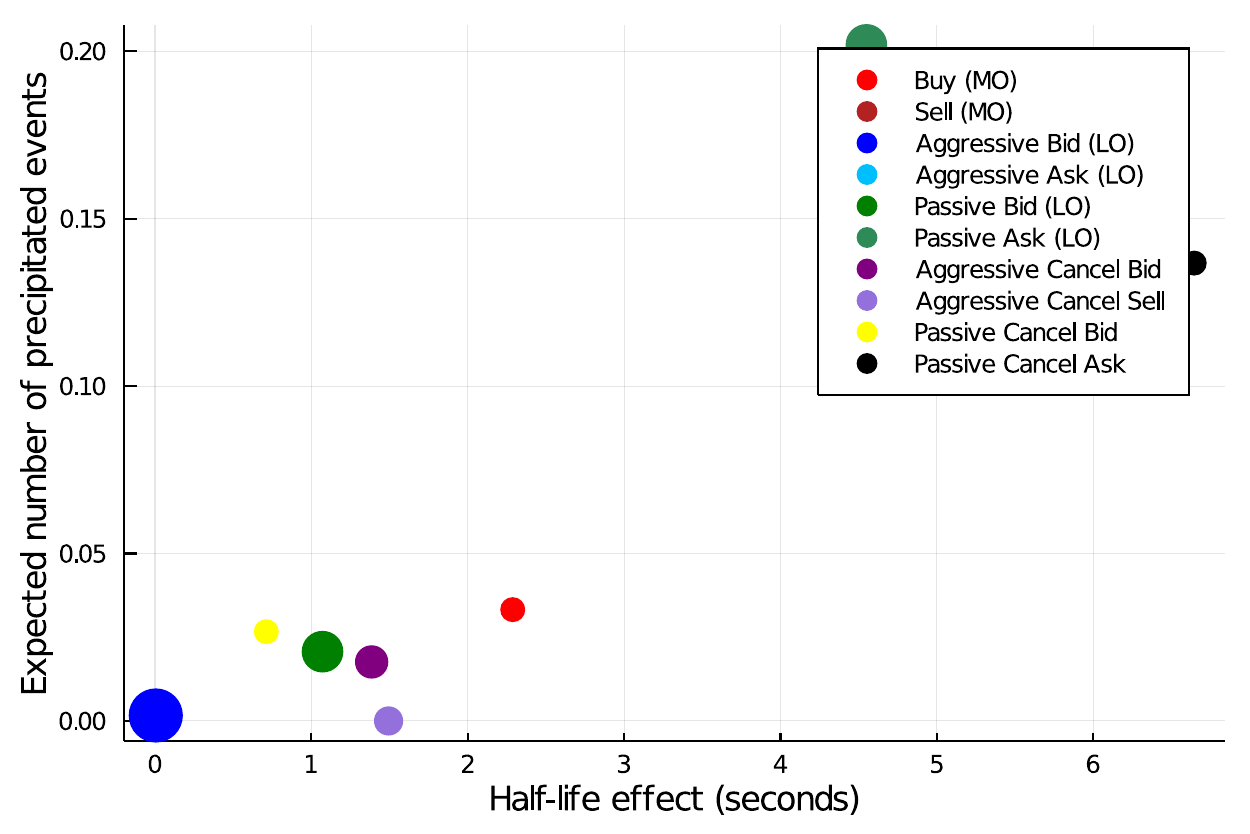}}
    \caption{Model 1 bubble plot for event types with large distortions. The size of the markers are proportional to the number of events of each type. The expected number of precipitated events are given by branching ratios of each type while the half-lives are calculated as in \cref{eq:half-life}. Aggressive cancels had the smallest half-life effect on sell MOs while passive bids had the greatest (figure \ref{subfig:if1-2}). Sell MOs had the greatest half-life effect on aggressive bids (figure \ref{subfig:if1-3}) while buy MOs had the greatest half-life effect on aggressive asks (figure \ref{subfig:if1-4}). Lastly, passive cancels of sell orders had the largest half life effect on itself with aggressive bids having the smallest (figure \ref{subfig:if1-10}).}
    \label{fig:impulse functions 1}
\end{figure*}

\begin{figure*}[htb]
    \centering
    \subfloat[Excitations on event type 1 \label{subfig:if2-1}
    ]{\includegraphics[width=.5\textwidth]{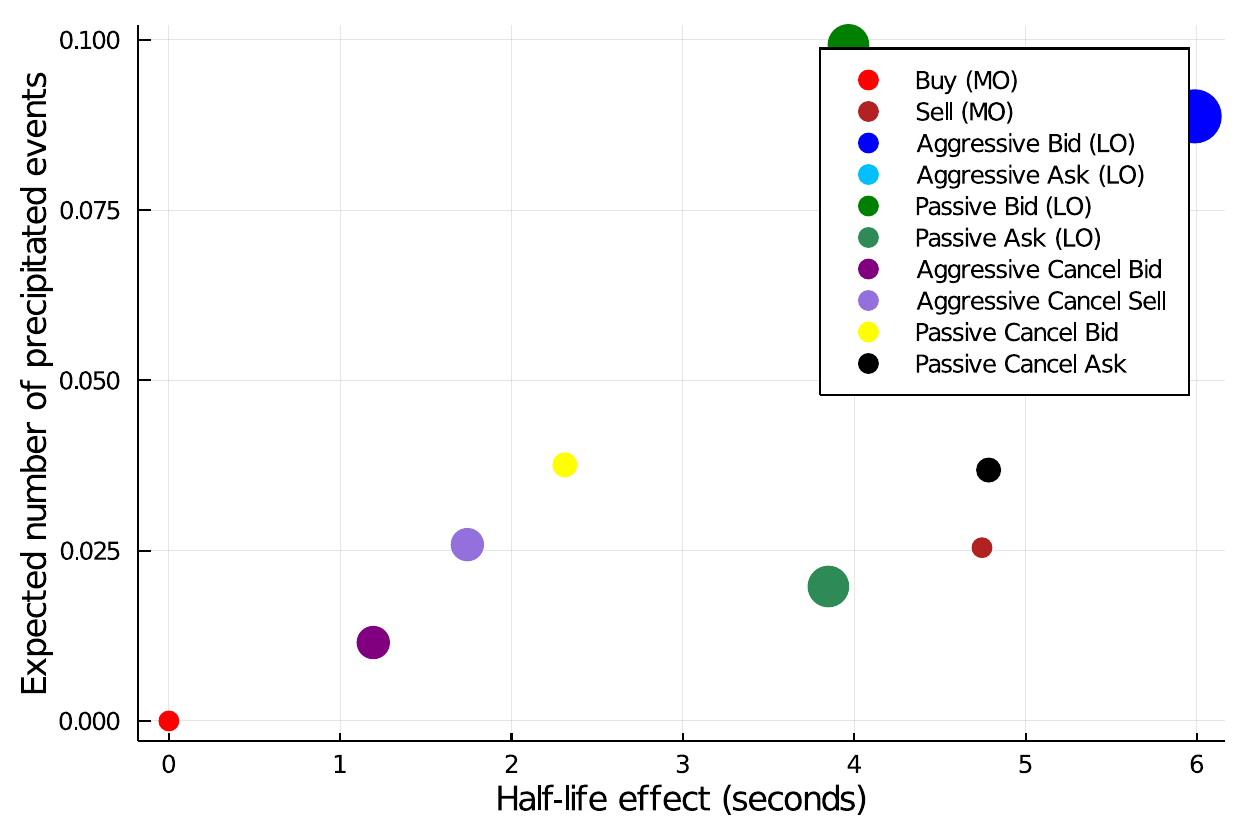}}
    \subfloat[Excitations on event type 2 \label{subfig:if2-2}
    ]{\includegraphics[width=.5\textwidth]{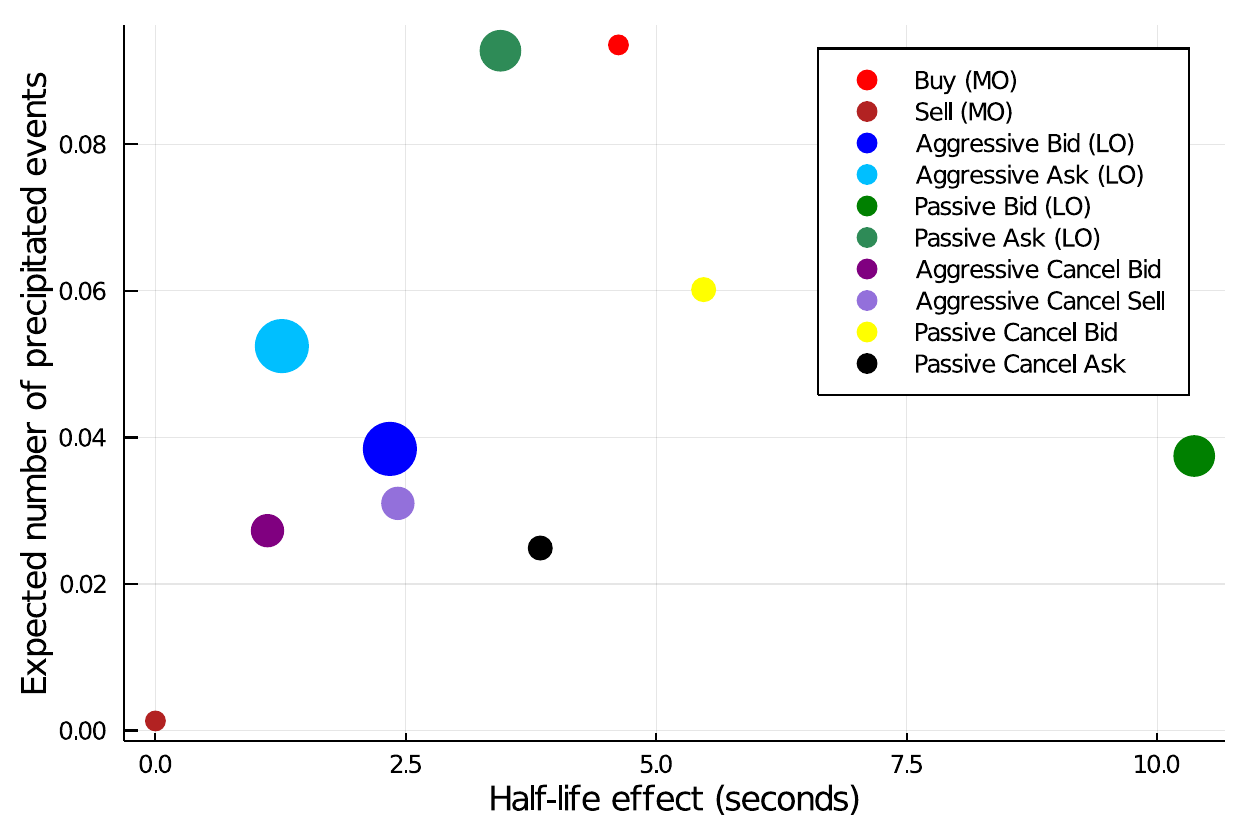}} \\
    \subfloat[Excitations on event type 9 \label{subfig:if2-9}
    ]{\includegraphics[width=.5\textwidth]{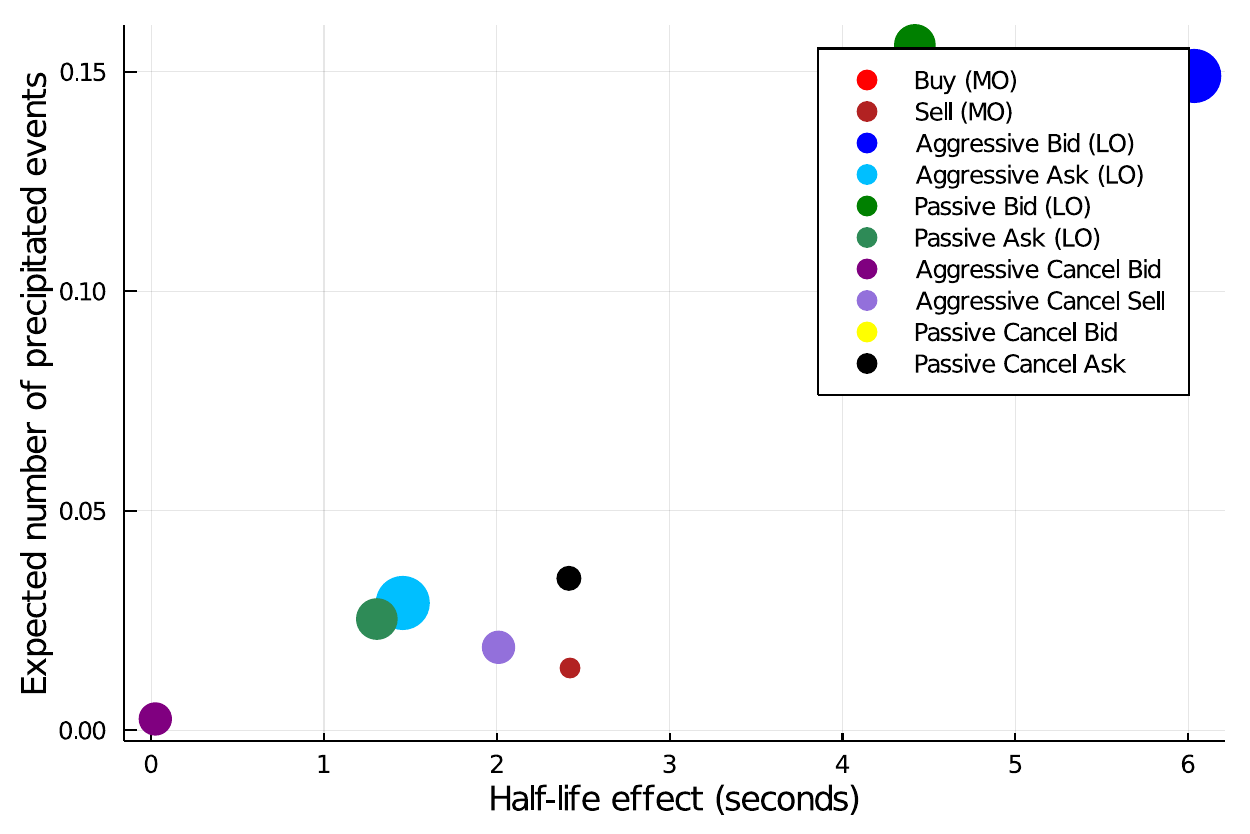}}
    \subfloat[Excitations on event type 10 \label{subfig:if2-10}
    ]{\includegraphics[width=.5\textwidth]{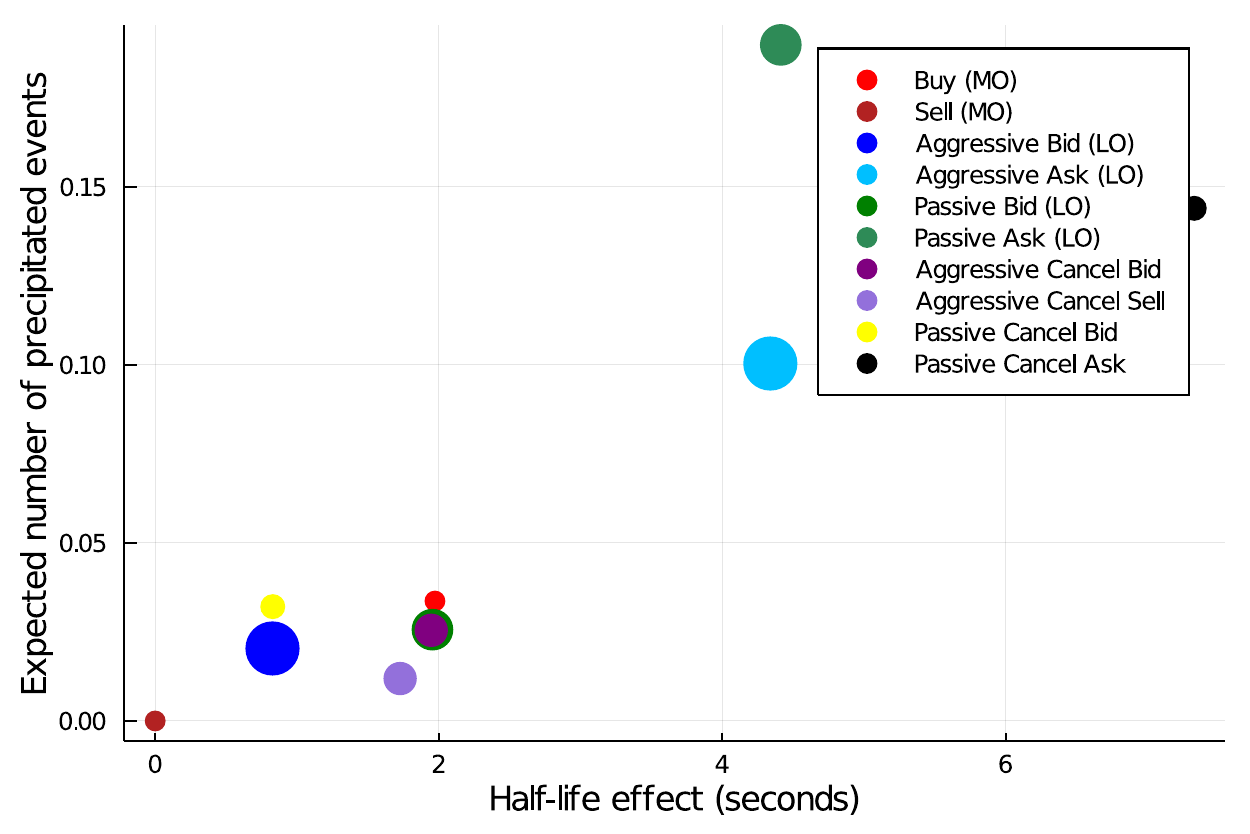}}
    \caption{Model 2 bubble plot for event types with large distortions. The size of the markers are proportional to the number of events of each type. The expected number of precipitated events are given by the branching ratios of each type while the half-lives are calculated as in \cref{eq:half-life}. Buy MOs had the smallest half-life effect on itself while aggressive bids had the greatest (figure \ref{subfig:if2-1}). Similarly, sell MOs had the smallest half-life effect on itself while passive bids had the greatest (figure \ref{subfig:if2-2}). Aggressive cancels of bids had the smallest half-life-effect on passive cancels of bids with aggressive bids having the greatest (figure \ref{subfig:if2-9}). Lastly, passive cancels of sell orders had the largest half life effect on itself with buy MOs having the smallest (figure \ref{subfig:if2-10}).}
    \label{fig:impulse functions 2}
\end{figure*}

Finally, figures \ref{fig:impulse functions 1} and \ref{fig:impulse functions 2} investigates the dynamics of individual excitation effects on event types with some severely distorted parameters found in figures \ref{fig:alpha distortion} and \ref{fig:beta distortion}. Each marker represents an individual effect with its branching ratio measured on the y-axis and its half-life\footnote{The half-life is the amount of time it takes for a given intensity effect $\alpha^{mn}$ to reduce by a half} (seconds) measured on the x-axis. The diameter of the markers are proportional to the excitation effect's event count. To obtain the half-life of a given intensity effect $\alpha^{mn}$ we need to solve
\begin{equation} \label{eq:half-life}
\begin{aligned}
\exp\left( - {t_{_{\sfrac{1}{2}}} \beta^{mn}} \right) = \sfrac{1}{2} ~ \text{ to get } ~ t_{_{\sfrac{1}{2}}} = \log(2) / \beta^{mn}.
\end{aligned}
\end{equation}
Branching ratios have the same interpretation as before, it is the average number of events of type $m$ caused by a single event of type $n$. The half-lives give us a sense of how long a particular excitation will last for. Moving through each figure one gets a sense for how each event's intensities are affected by all the others. 

\subsection{Parameter uncertainty \label{subsec:param uncertainty}}

We attempt to obtain confidence intervals for our parameters obtained through MLE using the result that the Fisher information can approximate the covariance matrix of the estimates.

The \textit{score} function is the first derivative of the log-likelihood given as
\begin{equation}
    S(\theta) = \frac{\partial}{\partial \theta} \ln \mathcal{L} (\theta).
\end{equation}
From Theorem 4 of \citet{Ogata1978}, $\frac{1}{\sqrt{T}} S\left( \theta \right)$ converges in law to $\mathcal{N}\left(0, \mathcal{I}\left( \theta \right) \right)$ as $T \rightarrow \infty$, where $\mathcal{I}\left( \theta \right)$ is the Fisher information. The Fisher information is therefore the variance of the score. Moreover, from Theorem 1 of \citet{Ogata1978}, the Fisher information can be computed as
\begin{equation}\label{eq:app2}
    \left[ \mathcal{I}\left( \theta \right) \right]_{i,j} = \mathbb{E}\left[ \frac{\partial \ln \mathcal{L} (\theta)}{\partial \theta_i} \frac{\partial \ln \mathcal{L} (\theta)}{\partial \theta_j} \right] = - \mathbb{E} \left[ \frac{\partial^2 \ln \mathcal{L} (\theta)}{\partial \theta_i \partial \theta_j} \right], \quad i,j = 1, \ldots 210.
\end{equation}

Instead of computing the Fisher information, we use the observed Fisher information given as
\begin{equation}
    \left[ I\left( \theta \right) \right]_{i,j} = - \frac{\partial^2 \ln \mathcal{L} (\theta)}{\partial \theta_i \partial \theta_j}.
\end{equation}
We make this choice because \cref{eq:app2} is usually computed analytically or using Monte Carlo methods which is not feasible for the size of our problem. Additionally, \citet{EH1978} favour the use of the observed Fisher information over the Fisher information. Therefore, our observed Fisher information is simply the negative Hessian of the log-likelihood evaluated at the maximum likelihood estimate which we compute numerically using the package ``ForwardDiff'' \cite{RLP2016}. Combining the results, we have a 95\% asymptotic confidence interval for each $\theta_i$ given by
\begin{equation}\label{eq:B5}
    \hat{\theta}_i \pm 1.96 \sqrt{I^{-1}_{i,i} \left( \hat{\theta} \right) / T},
\end{equation}
subject to a variety of conditions. This is the same approach \citet{L2007} used when constructing confidence intervals.

Interestingly, we found that 5 and 4 of our variance estimates were negative for model 1 and model 2 respectively. This was not a problem for the reference model. \citet{Freedman2007} mentioned that the observed information can generate negative variances estimates when maximum likelihood estimates are obtained on a restricted parameter space imposed by the null hypothesis. However, our calibration routine was performed on the unrestricted space and our validation indeed points towards the fact that we have obtained appropriate parameters for the respective data. \citet{MPR2007} demonstrated something interesting, which is that for many datasets, the observed information can lead to negative variance estimates as pointed out by \citet{VM2007}.

Looking further into this issue we found that the cause for our negative variance estimates was because the observed information was not positive definite. We had small negative eigenvalues. This result goes against Theorem 3 of \citet{Ogata1978} where our observed information should have been positive definite. As of now, it is unclear which assumptions from the Theorem are violated. However, we suspect that the issue may be numerical errors or we have insufficient observations for certain event types (due to events being dropped) leading to a bad likelihood. It is for this reason that we do not consider hypothesis tests using the Score or Wald test. Since our observed Fisher information generates negative variances, the test will be inconsistent as pointed out by \citet{Freedman2007}.

\begin{figure*}[htb]
    \centering
    \subfloat[Hawkes reference model]{\includegraphics[width=.33\textwidth]{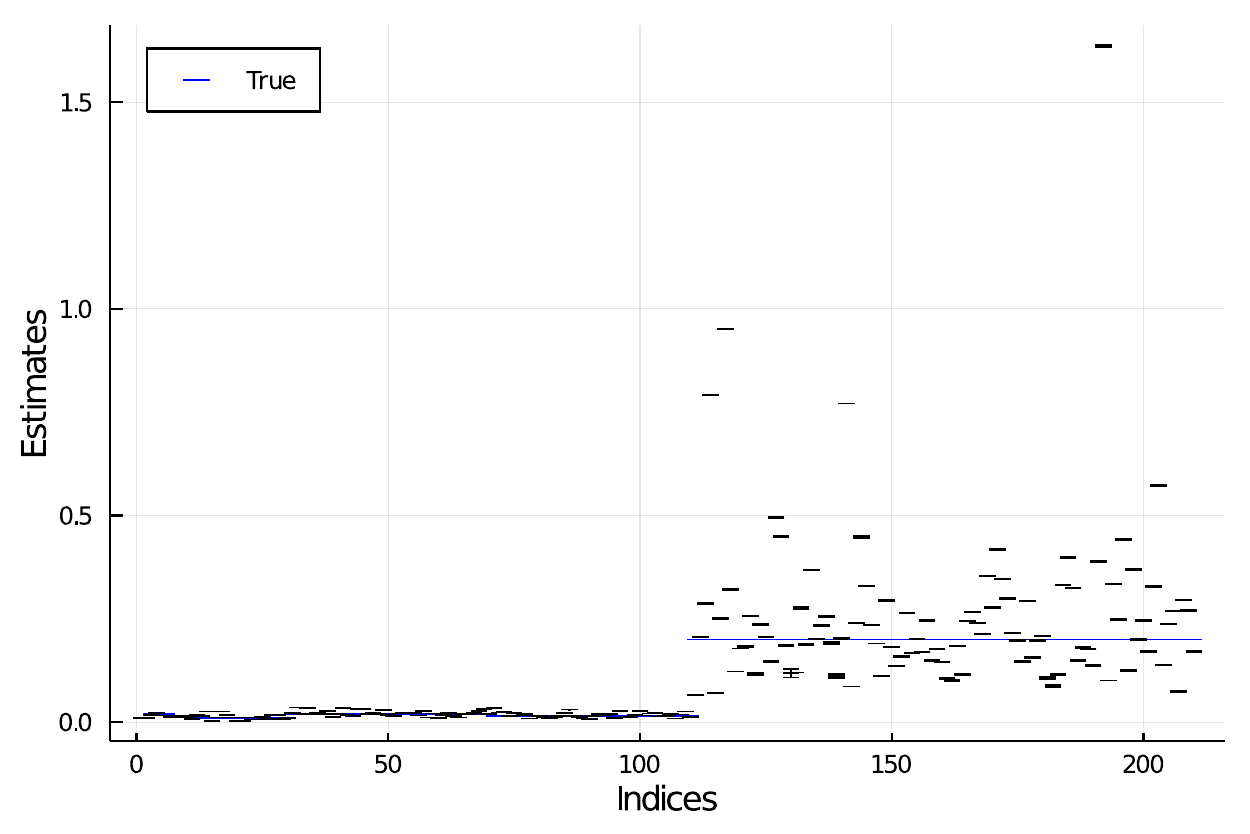}}
    \subfloat[Model 1]{\includegraphics[width=.33\textwidth]{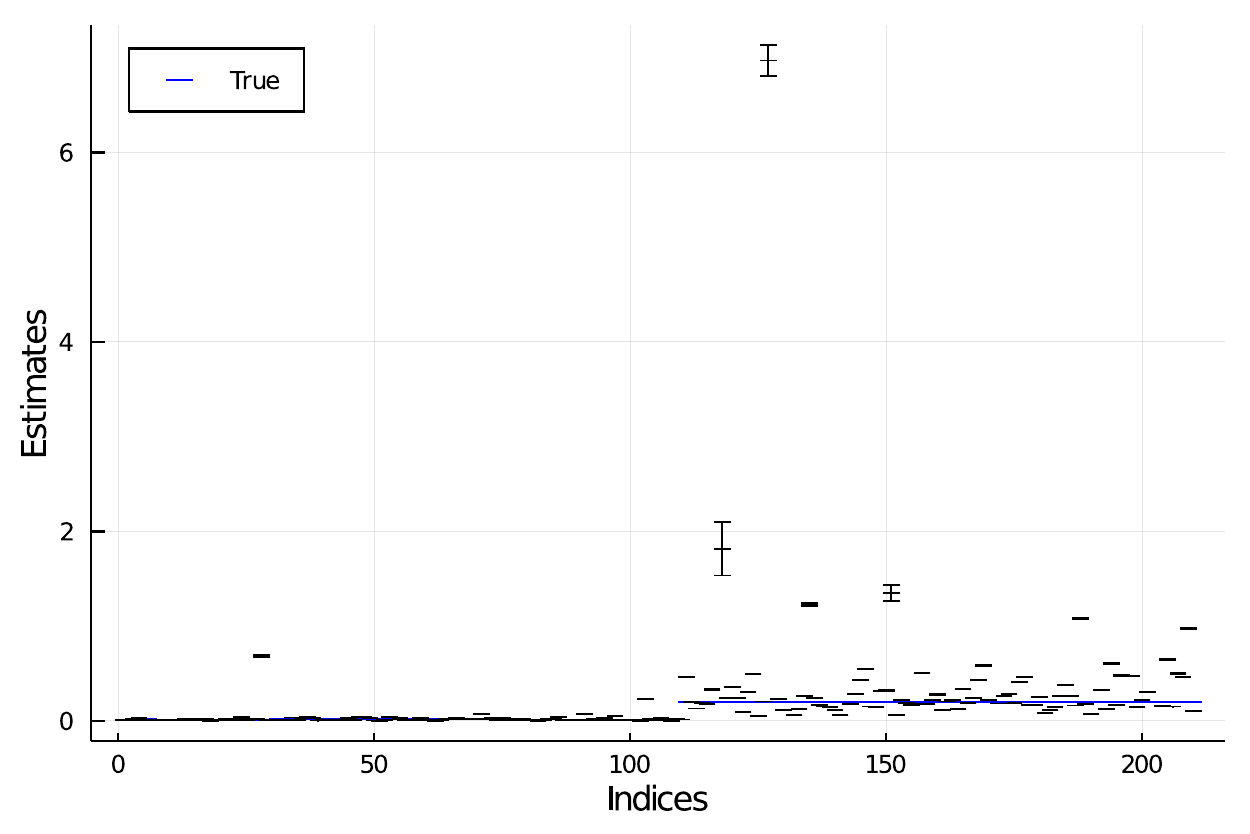}}
    \subfloat[Model 2]{\includegraphics[width=.33\textwidth]{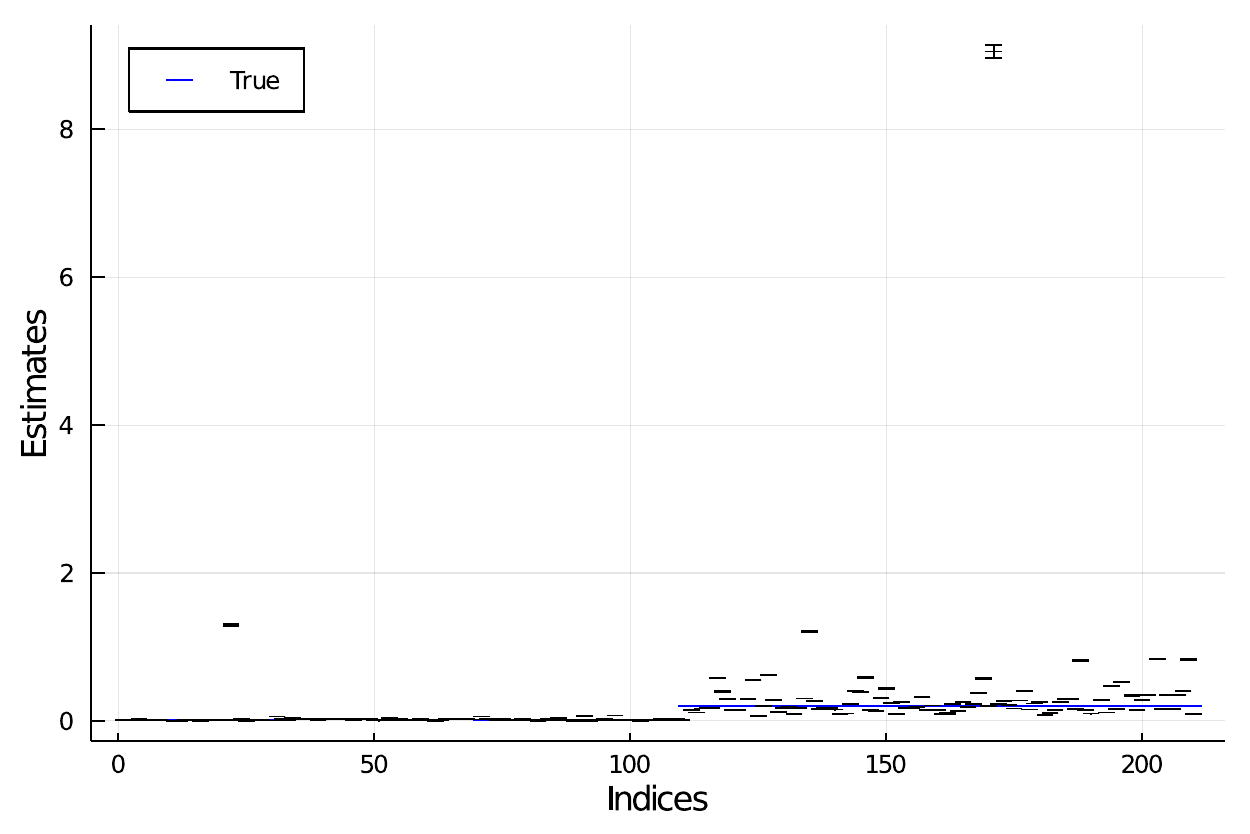}}
    \caption{Confidence intervals for parameter estimates for the three models computed from eq. \eqref{eq:B5}. Outliers are removed for improved readability with the largest parameter values being those that were distorted the most in figures \ref{fig:alpha distortion} and \ref{fig:beta distortion}. The top most data parameters from the reference model, model 1 and model 2 are $1.637 \pm 0.000109$, $6.972 \pm 0.00119$ and $9.049 \pm 0.00154$ respectively}
    \label{fig:confidence intervals}
\end{figure*}

Nonetheless, we visualise the estimated parameters along with their confidence intervals from \cref{eq:B5} in figure \ref{fig:confidence intervals}. To this end, we plot the estimated parameter and its confidence interval for 
$$
\hat{\theta} = \left( \hat{\mu}^1, \ldots, \hat{\mu}^{10}, \hat{\alpha}^{1,1}, \ldots, \hat{\alpha}^{1,10}, \hat{\alpha}^{2,1}, \ldots, \hat{\alpha}^{2,10}, \ldots, \hat{\alpha}^{10,1}, \ldots, \hat{\alpha}^{10,10}, \hat{\beta}^{1,1}, \ldots, \hat{\beta}^{1,10}, \hat{\beta}^{2,1}, \ldots, \hat{\beta}^{2,10}, \ldots, \hat{\beta}^{10,1}, \hat{\beta}^{10,10} \right).
$$
For parameters with a negative variance estimate, the confidence interval is excluded and only the estimate is visualised. Additionally, we exclude both the parameter value and confidence intervals for the parameters with extremely large deviation for better readability.

\begin{figure*}[htb]
    \centering
    \subfloat[Hawkes reference model]{\includegraphics[width=.33\textwidth]{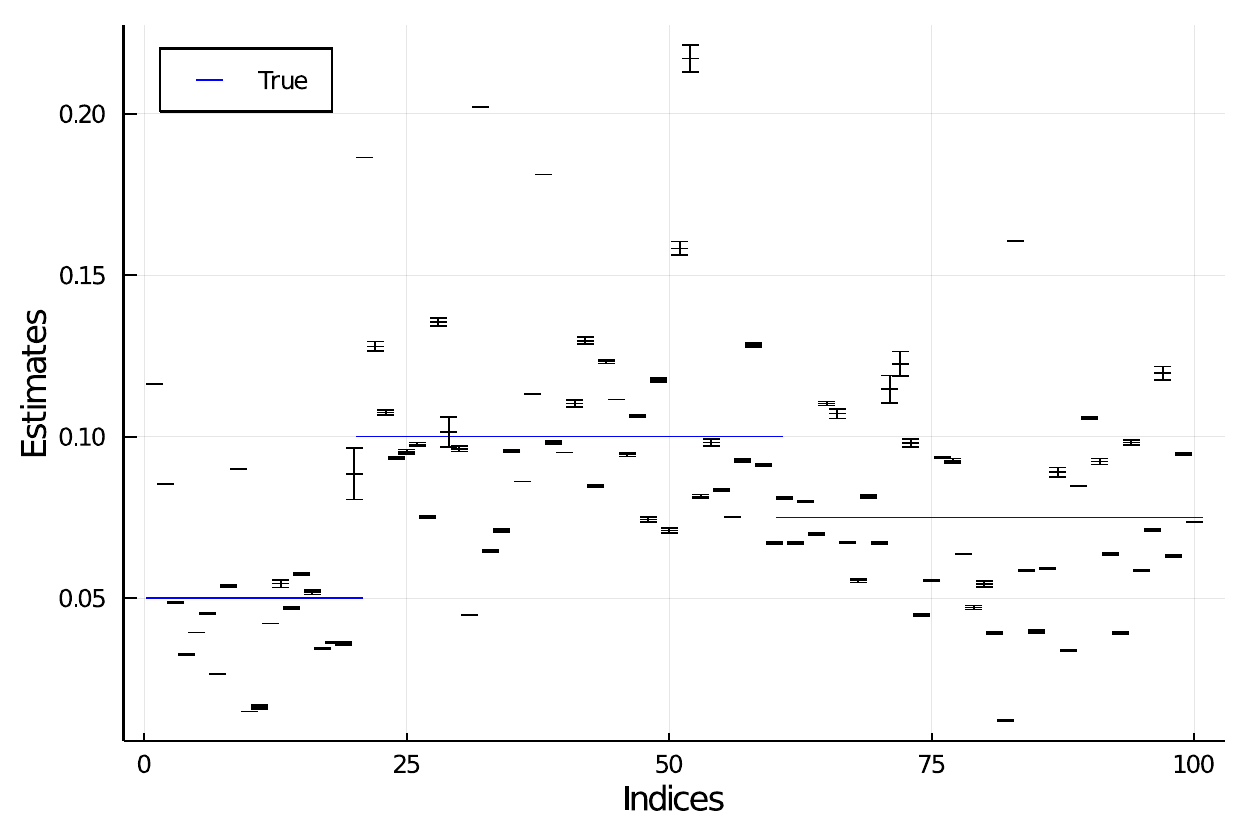}}
    \subfloat[Model 1]{\includegraphics[width=.33\textwidth]{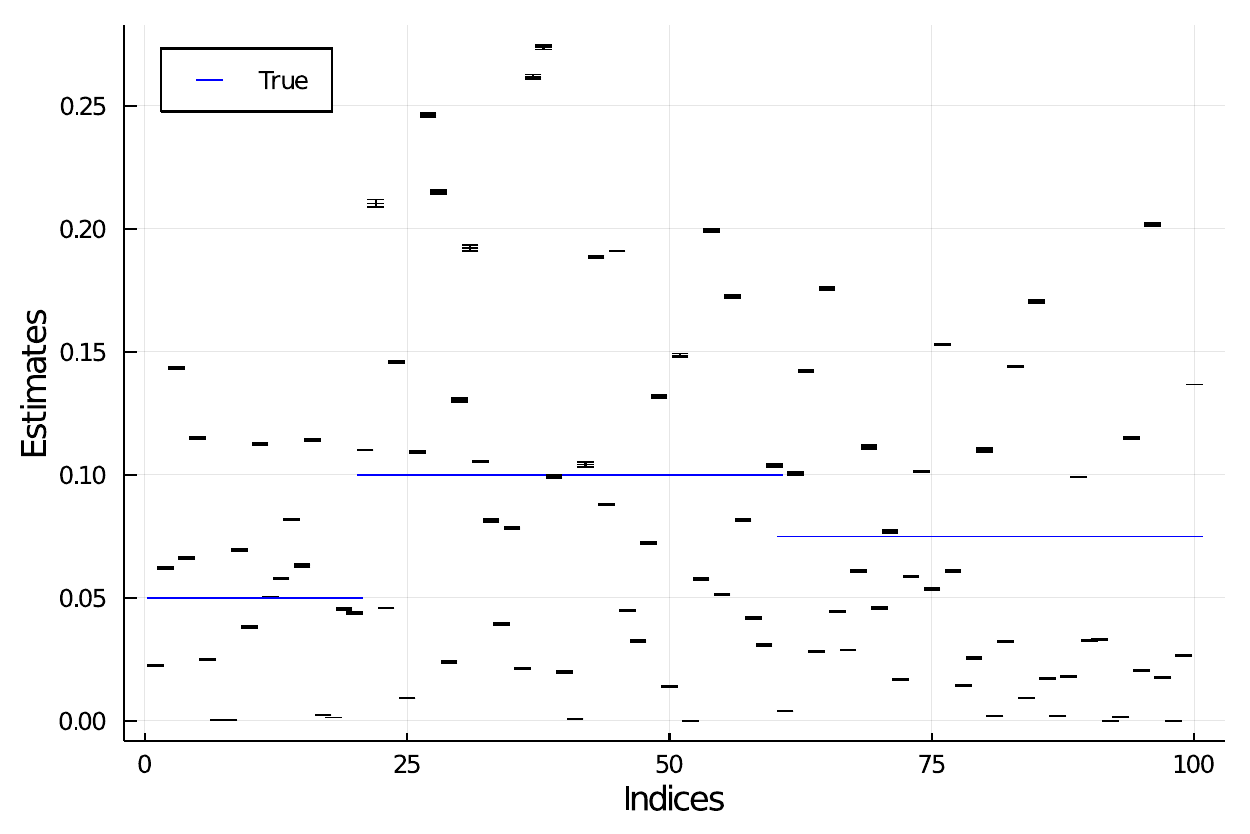}}
    \subfloat[Model 2]{\includegraphics[width=.33\textwidth]{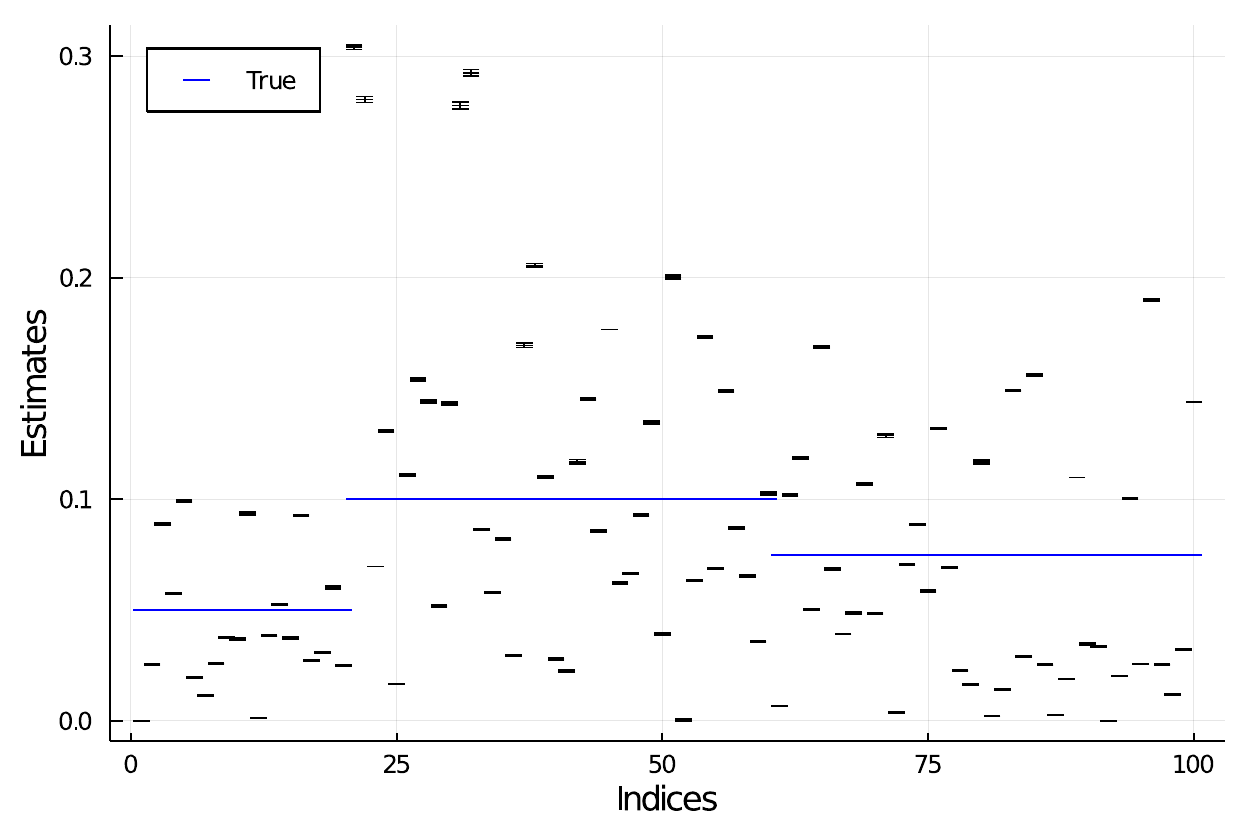}}
    \caption{Confidence intervals for branching ratios computed using the delta method for error propagation. Certain ratios exhibit larger intervals than others. Model 1 and 2 maintained a good balance of expected average event numbers but their branching ratios did not conform to that of the original process.}
    \label{fig:BRconfidence intervals}
\end{figure*}

We also include estimates and confidence intervals for the branching ratios. The variance for the branching ratio is obtained using the delta method for error propagation. We need to obtain the variance for $\Gamma^{nm} = f\left( \alpha^{nm}, \beta^{nm} \right) = \alpha^{nm}/\beta^{nm}$ but we only have variances and covariances estimates for $\alpha^{nm}$ and $\beta^{nm}$ (from the observed Fisher information). Therefore, we can approximate the variance of $\Gamma^{nm}$ as
$$
\begin{aligned}
\operatorname{Var} \Gamma^{nm}
&\approx \left(\frac{\partial f}{\partial \alpha^{nm}}\right)^{2} \operatorname{Var} \alpha^{nm} + \left(\frac{\partial f}{\partial \beta^{nm}}\right)^{2} \operatorname{Var} \beta^{nm} + \left(\frac{\partial f}{\partial \alpha^{nm}}\right)\left(\frac{\partial f}{\partial \beta^{nm}}\right) 2 \operatorname{Cov}(\alpha^{nm}, \beta^{nm}),
\end{aligned}
$$
where $\frac{\partial f}{\partial \alpha^{nm}} = 1/\hat{\beta}^{nm}$ and $\frac{\partial f}{\partial \beta^{nm}} = - \hat{\alpha}^{nm} / \left( \hat{\beta}^{nm} \right)^2$. The resulting estimates and confidence intervals can be found in figure \ref{fig:BRconfidence intervals}. The indices are ordered as
$$
\hat{\Gamma} = \left( \hat{\Gamma}^{1,1}, \ldots, \hat{\Gamma}^{1,10}, \hat{\Gamma}^{2,1}, \ldots, \hat{\Gamma}^{2,10}, \ldots, \hat{\Gamma}^{10,1}, \ldots, \hat{\Gamma}^{10,10} \right).
$$
Finally, confidence intervals are excluded for $\Gamma^{nm}$ which ended up with a negative variances or if $\alpha^{nm}$ and $\beta^{nm}$ had negative variances.

\end{document}

%% file: Preamble.tex
\usepackage{graphicx}
\graphicspath{{Figures/}}
\usepackage[caption=true]{subfig} % \subfloat
\usepackage{float} % [H]
\usepackage{booktabs} % \toprule
\usepackage{array}
\usepackage[colorlinks = true, allcolors = blue]{hyperref} % \href
\usepackage[linesnumbered, ruled, vlined]{algorithm2e}
\usepackage{multicol}
\usepackage{bm}

\usepackage{amssymb}
\usepackage{amsmath}
\usepackage{xfrac}
\usepackage{multirow}
\usepackage{cleveref}
\usepackage{bbm}
\usepackage{balance}

\newtheorem{theorem}{Theorem}

\makeatletter
\def\ps@pprintTitle{%
  \let\@oddhead\@empty
  \let\@evenhead\@empty
  \def\@oddfoot{\reset@font\hfil\thepage\hfil}
  \let\@evenfoot\@oddfoot
}
\makeatother